\documentclass{article}      

\usepackage[compact]{titlesec}

\DeclareRobustCommand\dotted{\tikz[baseline=-0.6ex]\draw[ultra  thick,dotted] (0,0)--(0.58,0);}

\DeclareRobustCommand\chain {\tikz[baseline=-0.6ex]\draw[very  thick,dash dot] (0,0)--(0.7,0);}
 
     \usepackage[final]{neurips_2021}

\usepackage{tikz}
\usepackage[utf8]{inputenc} 
\usepackage[T1]{fontenc}    
\usepackage{hyperref}       
\usepackage{url}            
\usepackage{booktabs}       
\usepackage{nicefrac}       
\usepackage{microtype}      
\usepackage{xcolor}         
\usepackage{setspace,url}
\usepackage{ifthen}
\usepackage{mathtools}
\usepackage{graphicx,epstopdf}
\usepackage{enumerate,comment}
\usepackage{enumitem}
\usepackage{color,xcolor,soul}
\usepackage{framed}
\usepackage{amsmath} 
\usepackage{amssymb}  
\usepackage{amsthm}
\usepackage{algorithm}
\usepackage{algpseudocode}

\newcounter{assumption}
\newcounter{subAssumption}
\newcounter{assumptionPrime}
\newcounter{subAssumptionPrime}
\renewcommand\theassumptionPrime{\arabic{assumptionPrime}'}
\renewcommand\thesubAssumption{\arabic{assumption}-\roman{subAssumption}}
\renewcommand\thesubAssumptionPrime{\arabic{assumptionPrime}'-\roman{subAssumptionPrime}}

\newenvironment{assumptions}{\refstepcounter{assumption}\setcounter{subAssumption}{0}\par
\rmfamily}

\newenvironment{subAssumption}{\refstepcounter{subAssumption}\par
   \noindent  \textbf{Assumption \thesubAssumption.} \em \rmfamily}

\newenvironment{assumption*}{\setcounter{assumptionPrime}{\theassumption-1}\refstepcounter{assumptionPrime}\par
   \noindent  \textbf{Assumption \theassumptionPrime.} \em \rmfamily}

\newenvironment{assumptions*}{\setcounter{assumptionPrime}{\theassumption-1}\refstepcounter{assumptionPrime}\setcounter{subAssumptionPrime}{0}\par
\rmfamily}

\newenvironment{subAssumption*}{\refstepcounter{subAssumptionPrime}\par
   \noindent  \textbf{Assumption \thesubAssumptionPrime.} \em \rmfamily}

\newcounter{example}
\newenvironment{example}{\refstepcounter{example}\par
   \noindent  \textbf{Example \theexample.} \em \rmfamily}

\newcounter{lemma}
\newenvironment{lemma}{\refstepcounter{lemma}\par
   \noindent  \textbf{Lemma \thelemma.} \em \rmfamily}

\newenvironment{lemma*}[1]{\par
   \noindent  \textbf{#1}. \em \rmfamily}

\newcounter{proposition}
\newenvironment{proposition}{\refstepcounter{proposition}\par
   \noindent  \textbf{Proposition \theproposition.} \em \rmfamily}

\newcounter{theorem}
\newenvironment{theorem}{\refstepcounter{theorem}\par
   \noindent  \textbf{Theorem \thetheorem.} \em \rmfamily}

\newcounter{corollary}
\newenvironment{corollary}{\refstepcounter{corollary}\par
   \noindent  \textbf{Corollary \thecorollary.} \em \rmfamily}

\newcounter{theoremApp}
\renewcommand\thetheoremApp{\Alph{theoremApp}}
\newenvironment{theorem*}{\refstepcounter{theoremApp}\par
   \noindent  \textbf{Theorem \thetheoremApp.} \em \rmfamily}

\newenvironment{myproof}{
   \noindent \textit{Proof:} \rmfamily}{\hfill $\square$}

\newcommand{\nn}{\nonumber}
\newcommand{\E}[1]{\mathbb{E}\left\{#1\right\}}
\newcommand{\inner}[2]{#1 \cdot #2}
\newcommand{\smooth}{\overline{\pi}}
\newcommand{\smoothFunc}{\overline{\mathrm{Br}}}
\newcommand{\argmax}{\mathop{\mathrm{argmax}}}
\newcommand{\val}{\mathrm{val}}
\newcommand{\Prob}[1]{\mathbb{P}\left\{#1\right\}}
\newcommand{\rhoa}{\rho_{\alpha}}
\newcommand{\rhob}{\rho_{\beta}}
\newcommand{\indicator}[1]{\mathbf{1}_{\left\{#1\right\}}}

\title{Decentralized $Q$-Learning in Zero-sum Markov Games}

\author{Muhammed O. Sayin\thanks{Equal contribution.}\\Bilkent University\\\texttt{sayin@ee.bilkent.edu.tr}  \And Kaiqing Zhang\footnotemark[1]\\MIT\\\texttt{kaiqing@mit.edu} \AND David S. Leslie\\Lancaster University\\ \texttt{d.leslie@lancaster.ac.uk} \And Tamer Ba\c{s}ar\\UIUC\\\texttt{basar1@illinois.edu} \And Asuman Ozdaglar\\MIT\\\texttt{asuman@mit.edu}}

\begin{document} 

\maketitle 

\begin{abstract}  
We study multi-agent  reinforcement learning (MARL) in infinite-horizon discounted zero-sum  Markov games. We focus on the practical but  challenging setting of  \emph{decentralized} MARL,  where agents make decisions without coordination   by a centralized  controller, but only based on their own payoffs and local actions executed.  The agents need not observe the opponent's actions or payoffs, possibly being even  oblivious to the presence of the opponent, nor be aware of the zero-sum structure of the underlying game, a setting also  referred to as \emph{radically uncoupled} in the literature of learning in games.   In this paper, we develop a radically uncoupled $Q$-learning dynamics that is both \emph{rational} and \emph{convergent}: the  learning dynamics converges to the best response to the opponent's strategy when the opponent follows an asymptotically stationary strategy;  when both agents adopt the learning dynamics, they converge to the Nash equilibrium of the game. The key challenge in this decentralized setting is the \emph{non-stationarity} of the environment from an agent's perspective, since {both her own payoffs and the system evolution depend on the actions of other agents,  and}  each agent adapts her policies simultaneously and independently.   To address this issue, we develop a two-timescale learning dynamics where each agent updates her local $Q$-function and value function estimates concurrently, with the latter happening at a slower timescale.  
\end{abstract}   

 
\section{Introduction}\label{sec:intro}

Reinforcement learning (RL) has achieved tremendous successes recently in a wide range of applications, including playing the game of Go \citep{silver2017mastering2},  playing video games (e.g., Atari \citep{mnih2015human} and Starcraft \citep{vinyals2019grandmaster}), robotics \citep{lillicrap2015continuous,kober2013reinforcement}, and autonomous driving \citep{shalev2016safe,sallab2017deep}. Most of these  applications  involve \emph{multiple} decision-makers, where the agents' rewards  and the evolution of the system are  affected by the joint behaviors  of all agents. This setting naturally leads to the problem of multi-agent RL (MARL). In fact, MARL is arguably one key ingredient of large-scale and reliable autonomy, and is significantly more challenging to analyze than single-agent RL. There has been a surging interest recently in both a deeper  theoretical and empirical understanding of  MARL; see  comprehensive overviews on this topic in \cite{busoniu2008comprehensive,zhang2019multi,hernandez2018multiagent}. 

The pioneering work that initiated the sub-area of MARL, where the model of {Markov/stochastic games} \citep{ref:Shapley53} has been considered  as a framework,  is \cite{littman1994markov}. Since then, there has been a plethora  of works on MARL in Markov games; see a detailed literature review in the supplementary material. 
These algorithms  can be broadly   categorized into two types: \emph{centralized/coordinated} and \emph{decentralized/independent} ones. For the former type, it is assumed that there exists a central controller for the agents, who can  access the agents joint actions and local observations. With full {awareness} of the game setup,  the central controller  coordinates the agents to optimize their own policies, and aims to compute an equilibrium.  
This centralized paradigm is typically suitable  for the scenarios when a simulator of the game is accessible \citep{silver2017mastering2,vinyals2017starcraft}. 
Most existing MARL algorithms in Markov games have focused on this paradigm.  

Nevertheless, in many practical multi-agent learning scenarios, e.g., multi-robot control \citep{wang2008machine}, urban traffic control \citep{kuyer2008multiagent}, as well as economics with rational decision-makers  \citep{fudenberg1998theory}, agents make decentralized decisions \emph{without} a coordinator.  Specifically, agents make updates independently with only \emph{local} observations of their own payoff and action histories, usually  in a {myopic} fashion. Besides its ubiquity in practice, this decentralized paradigm also has the advantage of being   \emph{scalable}, as each agent only cares about her own  policy and/or value functions, and the algorithms complexity do not suffer from the exponential dependence on the number of agents. 

Unfortunately,  establishing provably convergent decentralized  MARL algorithms is well-known to be  challenging; see the non-convergent cases (even in the fully cooperative setting) in \cite{tan1993multi,boutilier1996planning,claus1998dynamics}, and see \cite{matignon2012independent} for more empirical evidences. The key challenge is the \emph{non-stationarity} of the environment from each agent's perspective since all agents are adapting their policies simultaneously and independently. In other words, the opponent is not playing according to a stationary strategy.  This non-stationarity issue is in fact one of the core issues in  (decentralized)  MARL \citep{busoniu2008comprehensive,hernandez2017survey}. 

Studying when self-interested players can converge to an equilibrium through non-equilibrium adaptation is core question in the related literature of learning in games \citep{fudenberg1998theory,ref:Fudenberg09}. For example, simple and stylized learning dynamics, such as fictitious play, are shown to converge to an equilibrium in certain but important classes of games, e.g., zero-sum \citep{ref:Robinson51,ref:Harris98} and common-interest \citep{ref:Shapley96}, in repeated play of the same game.  
However, we cannot generalize these results to decentralized MARL {in \emph{Markov} games} (also known as stochastic games, introduced by \cite{ref:Shapley53}), because agents strategies affect not only the immediate reward, as in the repeated play of the same strategic-form game, but also the rewards that will be received in the future. Therefore, the configuration of the induced \emph{stage games} are not necessarily stationary in Markov games.  

\noindent\textbf{Contributions.} In this paper, we present a provably convergent decentralized MARL learning dynamics\footnote{To emphasize the difference from many existing MARL algorithms that focus on the \emph{computation} of Nash equilibrium, we refer to our update rule as \emph{learning dynamics}, following the literature of learning in games.}  for zero-sum discounted Markov games over an infinite horizon with minimal information available to agents. Particularly, each agent only has access to her immediate reward and the current state with perfect recall. They do not have access to the immediate reward the opponent receives. They do not know a model of their reward functions and the underlying state transitions probabilities. They are oblivious to the zero-sum structure of the underlying game. They also do not observe the opponent's actions. Indeed, they may even be oblivious to the presence of other agents. Learning dynamics with such minimal information is also referred to as being \emph{radically uncoupled} or \emph{value-based} in the literature of learning in games \citep{ref:Foster06,ref:Leslie05}.

To address the non-stationarity issue, we advocate a two-timescale adaptation of the individual $Q$-learning, introduced by \cite{ref:Leslie05} and originating in \cite{fudenberg1998theory}. Particularly, each agent infers the opponent's strategy indirectly through an estimate of the local $Q$-function (a function of the opponent's strategy) and simultaneously forms an estimate of the value function to infer the continuation payoff. The slow update of the value function estimate is natural since agents tend to change their strategies faster than their estimates (as observed in the evolutionary game theory literature, e.g., \cite{ref:Ely01,ref:Sandholm01}), but this also helps weakening the dependence between the configuration of the stage games (specifically the global $Q$-functions) and the strategies. We show the  almost sure convergence of the learning dynamics to the Nash equilibrium using the stochastic approximation theory,  by developing a novel Lyapunov function and identifying the sufficient conditions precisely later in \S\ref{sec:main}. Our techniques toward addressing these challenges might be of independent interest. We also  verify the convergence of the learning dynamics via numerical examples.

To the best of our knowledge, our learning dynamics  appears to be one of the first provably convergent decentralized MARL learning dynamics for Markov games that enjoy all the appealing properties below, addressing an important open question in the literature \citep{perolat2018actor,daskalakis2020independent}. In particular, our learning dynamics -- 
\begin{itemize}  [noitemsep,topsep=0pt,parsep=0pt,partopsep=0pt,leftmargin=25pt]
\item requires only minimal information available to the agents, i.e., it is a radically uncoupled learning dynamic, unlike many other MARL algorithm, e.g.,  \cite{perolat2015approximate,sidford2019solving,ref:Leslie20,ref:Sayin20,bai2020provable,xie2020learning,zhang2020model,liu2020sharp,shah2020reinforcement}; 
\item requires no coordination or communication between agents during learning. For example, agents always play the (smoothed) best response consistent with their self-interested decision-making, contrary to being coordinated to keep playing the same strategy within certain time intervals as in \cite{ref:Arslan17} and \cite{wei2021last};   
\item requires no  \emph{asymmetric} update rules and/or stepsizes for the agents unlike existing literature \citep{ref:Vrieze82}, \citep{ref:Bowling02}, \citep{ref:Leslie03}, \citep{daskalakis2020independent}, \citep{zhao2021provably,guo2021decentralized}. Such an asymmetry implies implicit coordination between agents to decide who follows which update rule or who chooses which stepsize  (and correspondingly who reacts fast or slow). \cite{daskalakis2020independent} refers to each agent playing a symmetric role in learning as \emph{strongly independent} learning. 
\item is both \emph{rational} and \emph{convergent}, a desired property for MARL (independent of whether it is centralized or decentralized), e.g., see \cite{bowling2001rational,busoniu2008comprehensive}. A MARL algorithm is rational if each agent can converge to best-response, when the opponent plays an (asymptotically) stationary strategy; and it can converge only to an equilibrium when all agents adopt it.  
\end{itemize}

A detailed literature review is deferred to the supplementary material due to space limitations. Of particular relevance are two recent works \cite{tian2020provably} and \cite{wei2021last} studied decentralized setting similar to ours. \cite{tian2020provably} focused on the exploration aspect for finite-horizon settings, and focused on minimizing a weak notion of regret without providing convergence guarantees under self-play.\footnote{Note that the same update rule with different stepsize and bonus choices and a  certified policy technique, however, can return a non-Markovian approximate Nash equilibrium policy pair in  the self-play setting; see \cite{bai2020near} for more details.} \cite{wei2021last} presented an optimistic variant of the gradient descent-ascent method that shares similar desired properties with our learning dynamics, with a strong guarantee of last-iterate convergence rates. However, the algorithm is delicately designed and different from the common value/policy-based RL update rules, e.g., $Q$-learning, as in our work. Moreover, to characterize finite-time convergence, in the model-free setting, the agents need to coordinate to interact multiple steps at each iteration of the algorithm, while our learning dynamics is coordination-free with natural update rules. These two works can thus be viewed as orthogonal to ours. After submitting our paper,  we became aware of a concurrent and independent work \cite{guo2021decentralized}, which also developed a decentralized algorithm for zero-sum Markov games with function approximation and finite-sample guarantees. In contrast to our learning dynamics, the algorithm requires a double-loop update rule,  and thus is asymmetric and requires coordination between agents.   The assumptions and technical novelties in both works  are also fundamentally different. See \S\ref{sec:dec_MARL_related_work} for a detailed comparison. 

\noindent\textbf{Organization.} The rest of the paper is organized as follows. We describe Markov games and our decentralized $Q$-learning dynamics in  \S\ref{sec:algorithm}. In  \S\ref{sec:main}, we present the assumptions and the convergence results. In  \S\ref{sec:simulation}, we provide numerical examples. We conclude the paper with some remarks in  \S\ref{sec:conclusion}. The supplementary material includes a detailed literature review and the proofs of technical results. 

\noindent\textbf{Notations.} Superscript denotes player identity. We represent the entries of vectors $q$ (or matrices $Q$) via $q[i]$ (or $Q[i,j]$). For two vectors $x$ and $y$, the inner product is denoted by $\inner{x}{y}=x^Ty$. For a finite set $A$, we denote the probability simplex over $A$ by $\Delta(A)$.


\section{Decentralized $Q$-learning in Zero-sum Markov Games} \label{sec:algorithm}

This section presents a decentralized $Q$-learning dynamics that does not need access to the opponent's actions and \emph{does not} need to know the zero-sum structure of the underlying Markov game. To this end, we first start by providing a formal description of Markov games.

Consider two players interacting with each other in a common dynamic environment, with totally conflicting objectives. The setting can be described by a two-player zero-sum Markov game, characterized by a tuple $\langle {S},\{{A}_s^i\}_{(i,s)\in\{1,2\}\times S},\{r_s^i\}_{(i,s)\in\{1,2\}\times S},p, \gamma \rangle$, where $S$ denotes the set of states, $A_s^i$ denotes the action set of player $i$ at state $s\in S$, and $\gamma\in[0,1)$ denotes the discount factor. At each interaction round, player $i$ receives a reward according to the function $r_s^i:A_s^1\times A_s^2 \rightarrow \mathbb{R}$. Since it is a zero-sum game, we have $r_s^1(a^1,a^2)+r_s^2(a^1,a^2) = 0$ for each joint action pair $(a^1,a^2)\in A_s^1\times A_s^2$. We denote the transition probability from state $s$ to state $s'$ given a joint action profile $(a^1_s,a^2_s)$ by $p(s'|s,a^1_s,a^2_s)$. 
Let us denote the stationary (Markov) strategy of player $i$ by $\pi^i := \{\pi_s^i \in \Delta(A_s^i)\}_{s\in S}$. We define the expected utility of player $i$ under the strategy profile $(\pi^1,\pi^2)$ as the expected discounted sum of the reward he collects over an infinite horizon
\begin{equation}
U^i(\pi^1,\pi^2) = \mathbb{E}_{\{a_k^j\sim \pi_{s_k}^j\}_{j\in\{1,2\}}}\left\{\sum_{k=0}^{\infty}\gamma^k r_{s_k}^i(a_k^1,a_k^2)\right\},
\end{equation}
where $\{s_0\sim p_o,s_{k+1}\sim p(\cdot|s_k,a_k^1,a_k^2),k\geq 0\}$ is a stochastic process describing the evolution of the state over time and $p_o\in\Delta(S)$ is the initial state distribution. The expectation is taken with respect to the {initial state}, randomness induced by state transitions and mixed strategies. 

A strategy profile $(\pi_*^1,\pi_*^2)$ is an $\varepsilon$-Nash equilibrium of the Markov game with $\varepsilon\geq 0$ provided that
\begin{subequations}
\begin{align}
&U^1(\pi_*^1,\pi_*^2) \geq U^1(\pi^1,\pi_*^2)-\varepsilon,\;\forall \pi^1\\
&U^2(\pi_*^1,\pi_*^2) \geq U^2(\pi_*^1,\pi^2)-\varepsilon,\;\forall \pi^2.
\end{align}
\end{subequations}
{A Nash equilibrium is an $\varepsilon$-Nash equilibrium with $\varepsilon=0$.} It is known that such a Nash equilibrium exists for discounted Markov games \citep{ref:Fink64,filar2012competitive}.

Given a strategy profile $\pi:=(\pi^1,\pi^2)$, we define the value function of player $i$ by
\begin{align}\label{eq:vpi}
v_{\pi}^i(s) = \mathbb{E}_{\{a_k^j\sim \pi_{s_k}^j\}_{j\in\{1,2\}}}\left\{r_s^i(a^1,a^2)+\gamma\sum_{s'\in S} v_{\pi}^i(s')p(s'|s,a^1,a^2)\right\}, \quad \forall s.
\end{align}
Note that $U^i(\pi^1,\pi^2)=\E{v_{\pi}^i(s_0)\,|\,s_0\sim p_o}$. We also define the $Q$-function that represents the value obtained for a given state and joint action pair as
\begin{align}\label{eq:Qpi}
Q_{\pi}^i(s,a^1,a^2) =\, &r_{s}^i(a^1,a^2)+ \gamma \sum_{s'\in S} v_{\pi}^i(s')p(s'|s,a^1,a^2), \quad \forall (s,a^1,a^2),
\end{align}
as well as the \emph{local} $Q$-function for player $i$ as
\begin{equation}\label{eq:q_def}
q^{i}_{\pi}(s,a^i) := \mathbb{E}_{a^{-i}\sim\pi_s^{-i}}\left\{Q_{\pi}^i(s,a^1,a^2)\right\},\quad \forall (s,a^i),
\end{equation}
where $-i$ denotes the opponent of player $i$.

{By the one-stage deviation principle, we can interpret the interaction between the players at each stage as they are playing an auxiliary stage game, in which the payoff functions are equal to the $Q$-functions, e.g., see \cite{ref:Shapley53}. However, the $Q$-functions, and correspondingly the payoff functions in these auxiliary stage games, change with evolving strategies of the players. Therefore, the plethora of existing results for repeated play of the same strategic-form game (e.g., see the review \cite{ref:Fudenberg09}) do not generalize here. To address this challenge, we next introduce our decentralized $Q$-learning dynamics.}

\begin{figure}[t]
\centering
\includegraphics[width=.73\textwidth]{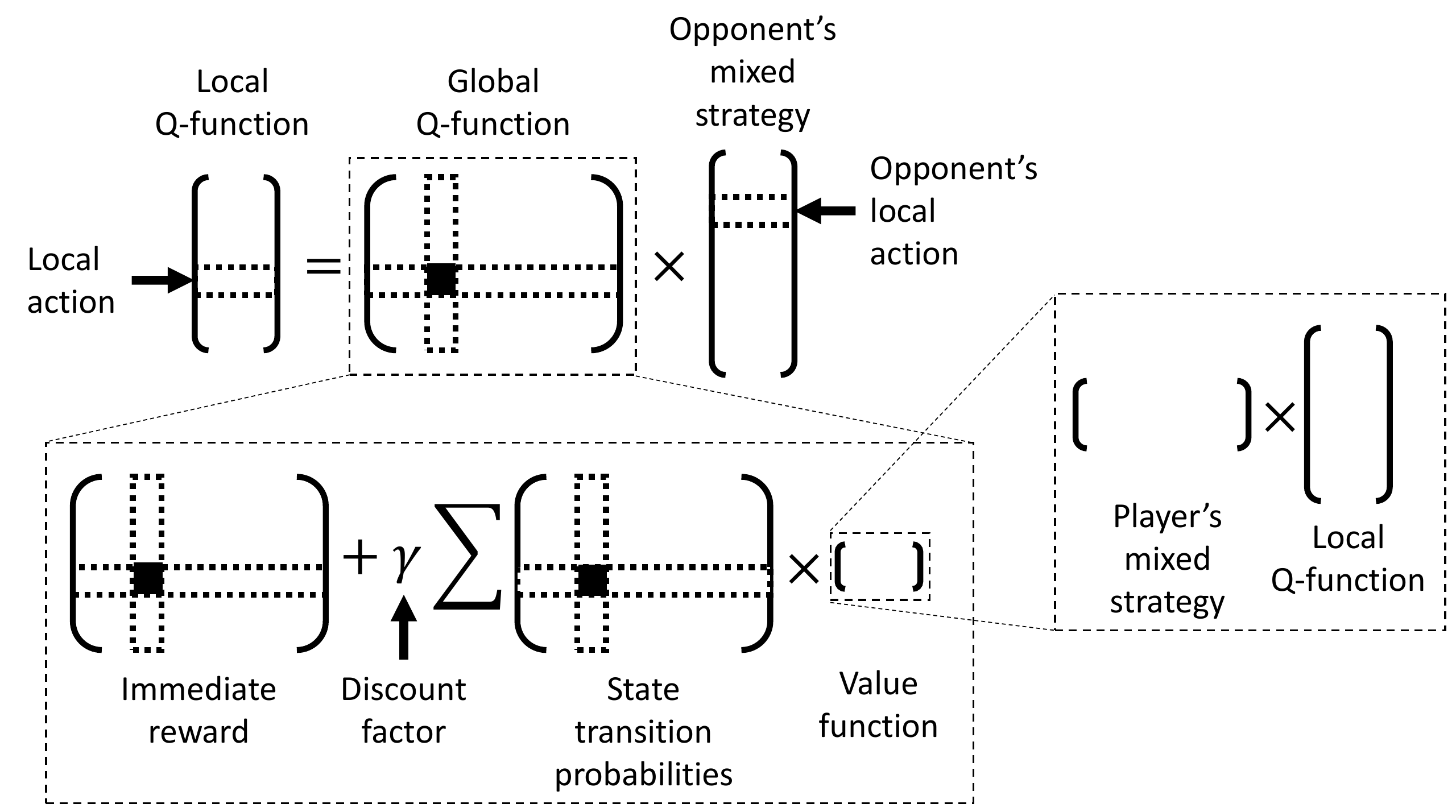}
\caption{A figurative illustration of the dependence between local $Q$-function, global $Q$-function, opponent's mixed strategy and the value function. From an agents' perspective, the local $Q$-function gives the expected values associated with local actions. Here, agents will infer the opponent's strategy over the local $Q$-function and estimate the value function at a slower timescale to ensure that the {\em implicit} global $Q$-function is relatively \emph{stationary} compared to the opponent's strategy.} \label{fig:model}
\vspace{-8pt}  
\end{figure}

\subsection{Decentralized $Q$-learning Dynamics}

In our decentralized $Q$-learning dynamics, minimal information is available to players. In other words, they only have access to the immediate reward received and current state visited with perfect recall. They do not observe the actions taken by the opponent.
Correspondingly, they cannot form a belief about the opponent's strategy based on the empirical play as in fictitious play \citep{fudenberg1998theory} or its  variant for stochastic games \citep{ref:Sayin20}. Instead, the players can look for inferring the opponent's strategy, e.g., by estimating the local $Q$-function since the local $Q$-function contains information about the opponent's strategy, as illustrated in Figure \ref{fig:model}. As seen in Figure \ref{fig:model}, however, the local $Q$-function also depends on the global $Q$-function while the global $Q$-function is not necessarily stationary since it depends on the value function, and therefore, depends on the players' evolving strategies. An estimate of the global $Q$ function which is slowly evolving would make it relatively stationary compared to the strategies. However, the players cannot estimate the global $Q$-function directly since they do not have access to the opponent's actions. Instead, they estimate the value function while updating it at a slower timescale. The slow update of the value function estimate makes the implicit global $Q$-function relatively stationary compared to the strategies. Therefore, the players can use the local $Q$-function estimate to infer the opponent's strategy.

Note that the local $Q$-function estimate for different actions would get updated asynchronously via the classical $Q$-learning algorithm since they would be updated only when the associated action is taken, however, the actions are likely to be taken at different frequencies. Instead, here the players update the local $Q$-function estimate via a  learning dynamics inspired from the individual $Q$-learning. The individual $Q$-learning, presented by \cite{ref:Leslie05} and originating in \cite{fudenberg1998theory}, is based on the $Q$-learning with soft-max exploration while the step sizes are normalized with the probability of the actions taken. This normalization ensures that the estimates for every action get updated at the same learning rate in the expectation. We elaborate further on this after we introduce the precise update of the local $Q$-function estimate later in this section.

Each player $i$ keeps track of $\{\hat{q}_{s,k}^i\}_{s\in S}$ and $\{\hat{v}_{s,k}^i\}_{s\in S}$ estimating, respectively, the local $Q$-function and the value function. Player $i$ updates $\{\hat{q}_{s,k}^i\}_{s\in S}$ at a faster timescale than $\{\hat{v}_{s,k}^i\}_{s\in S}$. Players also count the number of times each state $s$ is visited (until the current stage), denoted by $\# s$. 
 
We assume players know that the reward function takes values in $[-R,R]$ for some $R\in (0,\infty)$, i.e., $|r_s^i(a^1,a^2)|\leq R$ for all $(i,s,a^1,a^2)$. Therefore, player $i$ knows that his local $Q$-function $\|q_{\pi}^i(s,\cdot)\|_{\infty}\leq R/(1-\gamma)=: D$ and the value function $|v_{\pi}^i(s)|\leq D$ for any strategy profile $\pi$ and $s\in S$. Correspondingly, the players initiate these estimates arbitrarily such that $\|\hat{q}_{s,0}^i\|_{\infty} \leq D$ and $|\hat{v}_{s,0}^i|\leq D$, for all $s$. 

Let $s_k$ denote the current state at stage $k$. Player $i$ takes his action $a_{k}^i\in A^i_{s_k}$, drawn from the smoothed best response $\smoothFunc_{s_k}^i(\hat{q}_{s_k,k}^i,\tau_{\#s_k})\in \Delta(A_s^i)$, which depends on a temperature parameter $\tau_{\#s}>0$ associated with $\#s$. We define
\begin{equation}\label{eq:def_smooth_br}
\smoothFunc_s^i(q,\tau):=\argmax_{\mu\in\Delta(A_{s}^i)}~~\Big\{\inner{\mu}{q}+ \tau \nu_{s}^i(\mu)\Big\},
\end{equation}
{where $\nu^i_s$ is a smooth and strictly concave function whose gradient is unbounded at the boundary of the simplex $\Delta(A^i_s)$ \citep{fudenberg1998theory}.} The temperature parameter $\tau>0$ controls the amount of perturbation on the smoothed best response. The smooth perturbation ensures that there exists a unique maximizer $\smooth_{k}^i:=\smoothFunc_{s_k}^i(\hat{q}_{s_k,k}^i,\tau_{\#s_k})$. {Choosing $\nu^i_s(\mu):= -\sum_{a^i\in A_s^i}\mu[a^i] \log \mu[a^i]$ results in} the explicit characterization:
\begin{equation}\label{eq:def_smooth_br_2}
\smooth_{k}^i[a^i]=\frac{\exp\left(\hat{q}_{s_k,k}^i[a^i]/\tau_{\#s_k}\right)}{\sum_{\tilde{a}^i}\exp\left(\hat{q}_{s_k,k}^i[\tilde{a}^i]/\tau_{\#s_k}\right)}>0. 
\end{equation}

We let player $i$ update his local $Q$-function estimate's entry associated with the current state and local action pair $(s_k,a_{k}^i)$ towards the reward received plus the discounted continuation payoff estimate. To this end, we include $\hat{v}_{s_{k+1},k}^i$ as an unbiased estimate of the continuation payoff obtained by looking one stage ahead, as in the classical $Q$-learning introduced by \cite{ref:Watkins92}. Due to the one-stage look ahead, the update for the current state and local action can take place just after the game visits the next state. The update of $\hat{q}_{s_k,k}^i$ is given by
\begin{equation}\label{eq:q}
\hat{q}_{s_k,k+1}^i[a_{k}^i] = \hat{q}_{s_k,k}^i[a_{k}^i] + \bar{\alpha}_k^i\left(r_{k}^i + \gamma \hat{v}_{s_{k+1},k}^i - \hat{q}_{s_k,k}^i[a_{k}^i]\right),
\end{equation} 
where $\bar{\alpha}_k^i\in(0,1]$ is defined by $\bar{\alpha}_k^i := \min\left\{1,\frac{\alpha_{\#s_k}}{\smooth_{k}^i[a_{k}^i]}\right\}$, with $\{\alpha_{c}\}_{c>0}$ {a step size sequence}, and $r_{k}^i\in[-R,R]$ denotes the immediate reward of player $i$ at stage $k$. There is no update on others, i.e., $\hat{q}_{s,k+1}^i[a^i]=\hat{q}_{s,k}^i[a^i]$ for all $(s,a^i)\neq (s_k,a_k^i)$. {Inspired by the approach in \cite{ref:Leslie05}, the normalization addresses the asynchronous update of the entries of the local $Q$-function estimate and ensures that every entry of the local $Q$-function estimate is updated at the same rate in expectation.} We will show this explicitly in the proof of main theorem in the supplementary material. 
 
\begin{table}[t!]
\caption{Decentralized $Q$-learning dynamics in Markov games} 
\label{algo}
\hrule
\begin{algorithmic}[1]
\Require Keep track of $\{\hat{q}_{s,k}^i, \hat{v}_{s,k}^i,\#s\}_{s\in S}$.
\State {\bf Observe the current state} $s_k$.
\State {\bf Update} the entry of {\bf the local $Q$-function estimate} for the previous state $s_{k-1}$ and local action $a_{k-1}^i$ according to
$$ \hat{q}_{s_{k-1},k}^i[a_{k-1}^i] = \hat{q}_{s_{k-1},k-1}^i[a_{k-1}^i] + \bar{\alpha}_{k-1}^i\left(r_{k-1}^i + \gamma \hat{v}_{s_{k},k-1}^i - \hat{q}_{s_{k-1},k-1}^i[a_{k-1}^i]\right), $$
where $\bar{\alpha}_{k-1}^i = \min\left\{1,\frac{\alpha_{\#s_{k-1}}}{\smooth_{k-1}^i[a_{k-1}^i]}\right\}$, and $\hat{q}_{s,k}^i[a^i] = \hat{q}_{s,k-1}^i[a^i]$ for all $(s,a^i)\neq (s_{k-1},a_{k-1}^i)$.
\State {\bf Increment} the counter only for the current state by one.  
\State {\bf Take action} $a_{k}^i\sim \smooth_k^i$ where 
$$\smooth_{k}^i=\argmax_{\mu\in\Delta(A_{s_k}^i)}~~\Big\{\inner{\mu}{\hat{q}_{s_k,k}^i}+ \tau_{\#s_k} \nu_{s_k}^i(\mu)\Big\}.$$
\State {\bf Collect the local reward} $r_{k}^i$ and {\bf update the value function estimate} for the current state according to
$$\hat{v}_{s_k,k+1}^i = \hat{v}_{s_k,k}^i + \beta_{\#s_k}\left[\inner{\smooth_{k}^i}{\hat{q}_{s_k,k}^i} - \hat{v}_{s_k,k}^i\right]. $$
On the other hand, $\hat{v}_{s,k+1}^i = \hat{v}_{s,k}^i$ for all $s\neq s_{k}$.
\end{algorithmic}
\hrule
\vspace{-8pt} 
\end{table}

Simultaneous to updating the local $Q$-function estimate, player $i$ updates his value function estimate $\hat{v}_{s_k,k}^i$ towards $\inner{\smooth_{k}^i}{\hat{q}_{s_k,k}^i}$ corresponding to the expected value of the current state. However, the player uses a different step size $\{\beta_{c}\}_{c>0}$ and updates $\hat{v}_{s_k,k}^i$ according to 
\begin{equation}\label{eq:v}
\hat{v}_{s_k,k+1}^i = \hat{v}_{s_k,k}^i + \beta_{\#s_k}\left[\inner{\smooth_{k}^i}{\hat{q}_{s_k,k}^i} - \hat{v}_{s_k,k}^i\right].
\end{equation}
For other states $s\neq s_k$, there is no update on the value function estimate, i.e., $\hat{v}_{s,k+1}^i = \hat{v}_{s,k}^i$. To sum up, player $i$ follows the learning dynamics in Table \ref{algo}. {We emphasize that this dynamic is radically uncoupled since each player's update rule does not depend on the opponent's payoffs or actions.} In the next section, we study its convergence properties.


\section{Convergence Results}\label{sec:main}

We study whether the value function estimates in the learning dynamics, described in Table \ref{algo}, converge to an equilibrium value of the zero-sum Markov game. The answer is affirmative under certain conditions provided below precisely. {The first assumption (with two parts) is related to the step sizes and the temperature parameter, and not to the properties of the Markov game model.} 

\begin{assumptions}\label{assume:common}
\begin{subAssumption}\label{assume:common1}
The sequences $\{\alpha_c\in(0,1)\}_{c> 0}$ and $\{\beta_c\in(0,1)\}_{c> 0}$ are non-increasing and satisfy 
$\sum_{c=1}^{\infty}\alpha_c = \infty$, $\sum_{c=1}^{\infty}\beta_c = \infty$, and $\lim_{c\rightarrow\infty}\alpha_c=\lim_{c\rightarrow\infty}\beta_c = 0$.
\end{subAssumption}

\begin{subAssumption}\label{assume:common2}
Given any $M\in(0,1)$, there exists a non-decreasing polynomial function $C(\cdot)$ (which may depend on $M$) such that for any $\lambda\in(0,1)$ if $\left\{\ell\in\mathbb{Z}_+\,|\, \ell \leq c\mbox{ and }\frac{\beta_{\ell}}{\alpha_c} > \lambda\right\}\neq\varnothing$, then 
\begin{equation}
\max\left\{\ell\in\mathbb{Z}_+\,|\, \ell \leq c\quad\mbox{and}\quad\frac{\beta_{\ell}}{\alpha_c} > \lambda\right\} \leq Mc,\quad \forall c\geq C\left(\lambda^{-1}\right).
\end{equation}
\end{subAssumption}
\end{assumptions}

Assumption \ref{assume:common1} is a common  assumption used in stochastic approximation theory, e.g., see \cite{ref:Benaim99, ref:Borkar08}. On the other hand, Assumption \ref{assume:common2} imposes further condition on the step sizes than the usual two-timescale learning assumption, e.g., $\lim_{c\rightarrow\infty} \frac{\beta_c}{\alpha_c} = 0$, to address the asynchronous update of the iterates. Particularly, the iterates evolving at fast timescale can lag behind even the iterates evolving at slow timescale due to their asynchronous update. Assumption \ref{assume:common2} ensures that this can be tolerated when states are visited at comparable frequencies.   

For example, $\alpha_c = c^{-\rhoa}$ and $\beta_{c} = c^{-\rhob}$, where $0.5 < \rhoa < \rhob \leq 1$, satisfy Assumption \ref{assume:common} since it can be shown that there exists a non-decreasing polynomial, e.g., $C(x)= M_ox^{m}$, for all $x\geq 1$, where $M_o:=M^{-\rhob/(\rhob-\rhoa)}>0$ and $m\in\{m'\in\mathbb{Z}_+: m'\geq 1/(\rhob-\rhoa)\}$. We provide the relevant technical details in the supplementary material.

Such learning dynamics is not guaranteed to converge to an equilibrium in every class of zero-sum Markov games. For example, the underlying Markov chain may have an absorbing state such that once the game reaches that state, it stays there forever. Then, the players will not have a chance to improve their estimates for other states. Therefore, in the following, we identify two sets of assumptions (in addition to Assumption \ref{assume:common}) imposing increasingly stronger conditions on the underlying game while resulting in different convergence guarantees.   

\begin{assumptions}\label{assume:first}
\begin{subAssumption}\label{assume:firstIO}
Given any pair of states $(s,s')$, there exists at least one sequence of actions such that $s'$ is reachable from $s$ with some positive probability within a finite number, $n$, of stages.
\end{subAssumption}
\begin{subAssumption}\label{assume:first2}
The sequence $\{\tau_c\}_{c> 0}$ is non-increasing and satisfies $\lim_{c\rightarrow \infty} (\tau_{c+1}-\tau_c)/\alpha_c = 0$ and $\lim_{c\rightarrow\infty}\tau_c = \epsilon$ for some $\epsilon>0$. The step size $\{\alpha_c\}_{c>0}$ satisfies $\sum_{c=1}^{\infty} \alpha_c^{2}<\infty$. 
\end{subAssumption}
\end{assumptions}

In Assumption $\ref{assume:first}$, we do not let the temperature parameter go to zero. Next we let $\lim_{c\rightarrow\infty}\tau_c= 0$ but make the following assumption, imposing further condition on the underlying game {and $\{\alpha_c\}_{c>0}$} compared to Assumption \ref{assume:first} to ensure that each state gets visited infinitely often at comparable frequencies {and the normalization in the update of the local $Q$-function estimate does not cause an issue since it can be arbitrarily small when $\tau_c\rightarrow 0$}.

\begin{assumptions*}\label{assume:second}
\begin{subAssumption*}\label{assume:secondIO}
Given any pair of states $(s,s')$ and any infinite sequence of actions, $s'$ is reachable from $s$ with some positive probability within a finite number, $n$, of stages.
\end{subAssumption*} 
\begin{subAssumption*}\label{assume:second2}
The sequence $\{\tau_c\}_{c> 0}$ is non-increasing and satisfies $\lim_{c\rightarrow \infty} (\tau_{c+1}-\tau_c)/\alpha_c = 0$ and $\lim_{c\rightarrow\infty} \tau_c = 0$. The step size $\{\alpha_c\}_{c>0}$ satisfies
$\sum_{c=1}^{\infty} \alpha_c^{2-\rho}<\infty$, for some $\rho\in (0,1)$. There exists $C,C'\in(0,\infty)$ such that $\alpha_c^{\rho}\exp\left(4D/\tau_c\right)\leq C'$ for all $c\geq C$.  
\end{subAssumption*}
\end{assumptions*}

While being stronger than Assumption \ref{assume:firstIO},  Assumption \ref{assume:secondIO} is still weaker than those used in \cite{ref:Leslie20}. In \cite{ref:Leslie20}, it is assumed that there is a positive probability of reaching from any state to any other state in \emph{one} stage for any joint action taken by the players. {On the other hand, we say that a Markov game is {\em irreducible} if given any {\em pure} stationary strategy profile, the states visited form an irreducible Markov chain \citep{ref:Hoffman66,ref:Brafman02}. Assumption \ref{assume:secondIO} is akin to the {\em irreducibility assumption} for Markov games because the irreducibility assumption implies that there is a positive probability that any state is visited from any state within $|S|$ stages.} Furthermore, it reduces to the ergodicity property of Markov decision problems, e.g., see \cite{ref:Kearns02}, if one of the players has only one action at every state.  

As an example, $\alpha_c = c^{-\rhoa}$ and $\beta_{c} = c^{-\rhob}$, where $0.5 < \rhoa < \rhob \leq 1$ satisfies Assumptions \ref{assume:first2} and \ref{assume:second2}. There exists $\rho\in(0,2-1/\rhoa)$ for the latter since $\rhoa\in(0.5,1)$. To satisfy Assumption \ref{assume:first2}, the players can choose the temperature parameter $\{\tau_c\}_{c>0}$ 
as
\begin{equation}\label{equ:tau_to_epsilon}
\tau_c = \frac{1}{c}\bar{\tau}+\left(1-\frac{1}{c}\right)\epsilon, \quad\forall c>0
\end{equation}
with some $\bar{\tau}>0$.
On the other hand, to satisfy Assumption \ref{assume:second2}, they can choose the temperature parameter $\{\tau_c'\}_{c>0}$ as
\begin{equation}\label{equ:tau_to_zero}
\tau_c' = \bar{\tau}\left(1+\bar{\tau}\frac{\rhoa\rho}{4D}\log(c)\right)^{-1}, \quad \forall c>0.
\end{equation}
Alternative to \eqref{equ:tau_to_epsilon}, $\tau_c=\max\{\epsilon,\tau_c'\}$ also satisfies Assumption \ref{assume:first2} while having similar nature with \eqref{equ:tau_to_zero}. We provide the relevant technical details in the supplementary material.

We have the following key properties for the estimate sequence  generated by our learning dynamics.

\begin{proposition}\label{prop:bound}
Since $\bar{\alpha}_{k}^i\in(0,1]$ for all $k\geq 0$, $\beta_c\in(0,1)$ for all $c>0$, $\|\hat{q}_{s,0}^i\|_{\infty}\leq D$ and $|\hat{v}_{s,0}^i|\leq D$ for all $(i,s)$, the iterates are bounded, i.e., $\|\hat{q}_{s,k}^i\|_{\infty}\leq D$ and $|\hat{v}_{s,k}^i|\leq D$ for all $(i,s)$ and $k\geq 0$.
\end{proposition}

\begin{proposition}\label{prop:reduce}
Suppose that Assumption \ref{assume:common1} and either Assumption \ref{assume:first2} or \ref{assume:second2} hold. Then, there exists $C_s\in\mathbb{Z}_+$ for each $s\in S$ such that $\alpha_c\exp\left(2D/\tau_c\right)< \min_{i=1,2}\{|A_s^i|^{-1}\}$, for all $c\geq C_s$. Correspondingly, the update of the local $Q$-function estimate \eqref{eq:q} reduces to
\begin{equation}\label{eq:q_sim}
\hat{q}_{s_k,k+1}^i[a_{k}^i] = \hat{q}_{s_k,k}^i[a_{k}^i] + \frac{\alpha_{\#s_k}}{\smooth_{k}^i[a_{k}^i]}\left(r_{k}^i + \gamma \hat{v}_{s_{k+1},k}^i - \hat{q}_{s_k,k}^i[a_{k}^i]\right),
\end{equation}
for all $\#s_k \geq C_{s_k}$ since $\alpha_{\#s_k}/\smooth_{k}^i[a_{k}^i]\leq |A_{s_k}^i|\alpha_{\#s_k}\exp\left(2D/\tau_{\#s_k}\right)\leq 1$ by \eqref{eq:def_smooth_br_2}.
\end{proposition}

\begin{proposition}\label{prop:io}
Suppose that either Assumption \ref{assume:first} or Assumption \ref{assume:second} holds. Then, at any stage $k$, there is a fixed positive probability, e.g., $\underline{p}>0$, that the game visits any state $s$ at least once within $n$-stages independent of how players play. Therefore, $\#s\rightarrow \infty$ as $k\rightarrow\infty$ with probability $1$. 
\end{proposition}

{Proposition \ref{prop:bound} says that defining $\bar{\alpha}_k^i= \min\{1,\alpha_{\#s_k}/\smooth_{k}^i[a_{k}^i]\}$ ensures that the iterates remain bounded. On the other hand, Propositions \ref{prop:reduce} and \ref{prop:io} say that the update of the local $Q$-function estimates reduces to \eqref{eq:q_sim} where $\bar{\alpha}_k^i =\alpha_{\#s_k}/\smooth_{k}^i[a_{k}^i]$ after a finite number of stages, almost surely.} The following theorem characterizes the convergence properties of the $Q$-learning dynamics presented.

\begin{theorem}\label{thm:main}
Suppose that both players follow the learning dynamics described in Table \ref{algo} and Assumption \ref{assume:common} holds. Let $v_{\pi_*}^i$ and $Q_{\pi_*}^i$, as described resp. in \eqref{eq:vpi} and \eqref{eq:Qpi}, be the unique values associated with some equilibrium profile $\pi_*=(\pi_*^1,\pi_*^2)$ of the underlying zero-sum Markov game. Then, the asymptotic behavior of the value function estimates $\{\hat{v}_{s,k}^i\}_{k\geq 0}$ is given by
\begin{subequations}\label{eq:result}
\begin{align}
&\limsup_{k\rightarrow\infty}|\hat{v}_{s,k}^i - v_{\pi_*}^i(s)| \leq \epsilon \xi g(\gamma),\quad \mbox{under Assumption \ref{assume:first}},\label{equ:result_to_eps}\\
&\lim_{k\rightarrow\infty}|\hat{v}_{s,k}^i - v_{\pi_*}^i(s)| = 0,\hspace{0.6in}\mbox{under Assumption \ref{assume:second}},\label{equ:result_to_zero}
\end{align}
\end{subequations}
for all $(i,s)\in\{1,2\}\times S$, with probability (w.p.) $1$, where $\xi:= \max_{s'\in S}\left\{\log(|A_{s'}^1||A_{s'}^2|)\right\}$, and $g(\gamma)=\frac{2+\lambda-\lambda\gamma}{(1-\lambda\gamma)(1-\gamma)}$ with some $\lambda\in(1,1/\gamma)$.

Furthermore, let $\hat{\pi}_{s,k}^i$ be the weighted time-average of the smoothed best response updated as
$$
\hat{\pi}_{s,k+1}^i=\hat{\pi}_{s,k}^i + \mathbf{1}_{\{s=s_k\}}\alpha_{\#s}\left(\smooth_k^i - \hat{\pi}_{s,k}^i\right).
$$
Then, the asymptotic behavior of these weighted averages $\{\hat{\pi}_{k}^i\}_{k\geq 0}$ is given by
\begin{subequations}\label{eq:result2}
\begin{align}
&\limsup_{k\rightarrow\infty}\Big(\max_{\pi^i}~v_{\pi^i,\hat{\pi}_{k}^{-i}}^i(s) - v_{\hat{\pi}_{k}^{i},\hat{\pi}_{k}^{-i}}^i(s) \Big)\leq \epsilon \xi h(\gamma),\quad \mbox{under Assumption \ref{assume:first}},\\
&\lim_{k\rightarrow\infty}\Big(\max_{\pi^i}~v_{\pi^i,\hat{\pi}_{k}^{-i}}^i(s) - v_{\hat{\pi}_{k}^{i},\hat{\pi}_{k}^{-i}}^i(s) \Big) = 0,\hspace{.6in}\mbox{under Assumption \ref{assume:second}},\end{align}
\end{subequations}
for all $(i,s)\in\{1,2\}\times S$, w.p. $1$, where $h(\gamma) = \big[4\gamma\cdot g(\gamma)+2(1+\lambda)/(1-\lambda\gamma)\big]/(1-\gamma)$, i.e., these weighted-average strategies  converge to near or exact equilibrium depending on whether Assumption \ref{assume:first} or \ref{assume:second} hold.
\end{theorem}

A brief sketch of the proof is as follows: We decouple the dynamics specific to a single state from others by addressing the asynchronous update of the local $Q$-function estimate and the diminishing temperature parameter. We then approximate the dynamics specific to a single state via its limiting ordinary differential equation (o.d.e.) as if the iterates evolving at the slow timescale are time-invariant. We present a novel Lyapunov function for the limiting o.d.e. to characterize the limit set of the discrete-time update. This Lyapunov function shows that the game perceived by the agents become zero-sum asymptotically and the local $Q$-function estimates are asymptotically belief-based. Finally, we use this limit set characterization to show the convergence of the dynamics across every state by using asynchronous stochastic approximation methods, e.g., see \cite{ref:Tsitsiklis94}. 

The following corollary to Theorem \ref{thm:main} highlights the {\em rationality} property of our learning dynamics.

\begin{corollary} \label{cor:rational}
Suppose that player $-i$ follows an (asymptotically) stationary strategy 
$\{\tilde{\pi}_s^{-i}\in \mathrm{int}\Delta(A_s^i)\}_{s\in S}$ while player $i$ adopts the learning dynamics described in Table \ref{algo}, and Assumption \ref{assume:common} holds. 
Then, the asymptotic behavior of the value function estimate $\{\hat{v}_{s,k}^i\}_{k\geq 0}$ is given by 
\begin{subequations}
\begin{align}
&\limsup_{k\rightarrow\infty}\left|\hat{v}_{s,k}^i - \max_{\pi^i}~v_{\pi^i,\tilde{\pi}^{-i}}^i(s)
\right| \leq \epsilon \xi^i g(\gamma),\quad \mbox{under Assumption \ref{assume:first}},\\
&\lim_{k\rightarrow\infty} \left|\hat{v}_{s,k}^i - \max_{\pi^i}~v_{\pi^i,\tilde{\pi}^{-i}}^i(s)
\right| = 0,\hspace{.5in}\quad \mbox{under Assumption \ref{assume:second}}
\end{align}
\end{subequations}
for all $s\in S$, w.p. $1$, where $\xi^i:= \max_{s'\in S}\left\{\log(|A_{s'}^i|)\right\}$ and $g(\cdot)$ is as described in Theorem \ref{thm:main}. 

Furthermore, the asymptotic behavior of the weighted averages $\{\hat{\pi}_{k}^i\}_{k\geq 0}$, described in Theorem \ref{thm:main}, is given by
\begin{subequations}\label{eq:result2}
\begin{align}
&\limsup_{k\rightarrow\infty}\Big(\max_{\pi^i}~v_{\pi^i,\tilde{\pi}^{-i}}^i(s) - v_{\hat{\pi}_{k}^{i},\tilde{\pi}^{-i}}^i(s)
\Big)
\leq \epsilon \xi^i h(\gamma),\quad\hspace{-.05in} \mbox{under Assumption \ref{assume:first}},\\
&\lim_{k\rightarrow\infty}\Big(\max_{\pi^i}~v_{\pi^i,\tilde{\pi}^{-i}}^i(s) - v_{\hat{\pi}_{k}^{i},\tilde{\pi}^{-i}}^i(s)\Big)
= 0,\hspace{.58in}\mbox{under Assumption \ref{assume:second}},\end{align}
\end{subequations}
for all $s\in S$, w.p. $1$, where $h(\gamma)$ is as described in Theorem \ref{thm:main}, i.e., these weighted-average strategies  converge to near or exact best-response strategy, depending on whether Assumption \ref{assume:first} or \ref{assume:second} hold.
\end{corollary}


\section{Simulation Results} \label{sec:simulation}

\begin{figure*}[t]
	\centering
	\hspace{-20pt}
	\begin{minipage}[c]{0.5\linewidth}
		\includegraphics[width=\linewidth]{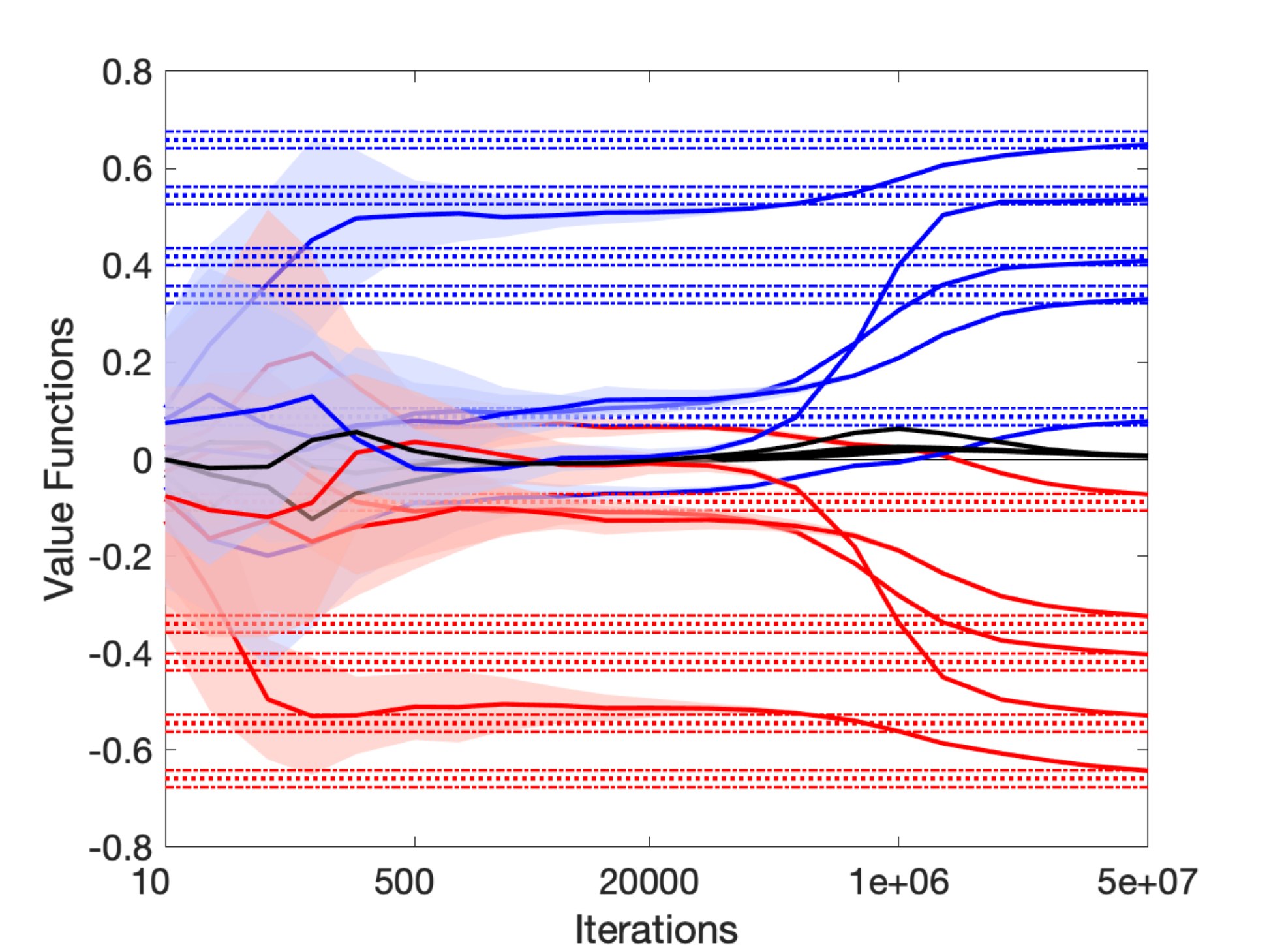}
		 \centering{\small(a) Convergence under Assumption \ref{assume:first}}
	\end{minipage} 
	\hspace{-19pt}
	\begin{minipage}[c]{0.5\linewidth}
		\includegraphics[width=\linewidth]{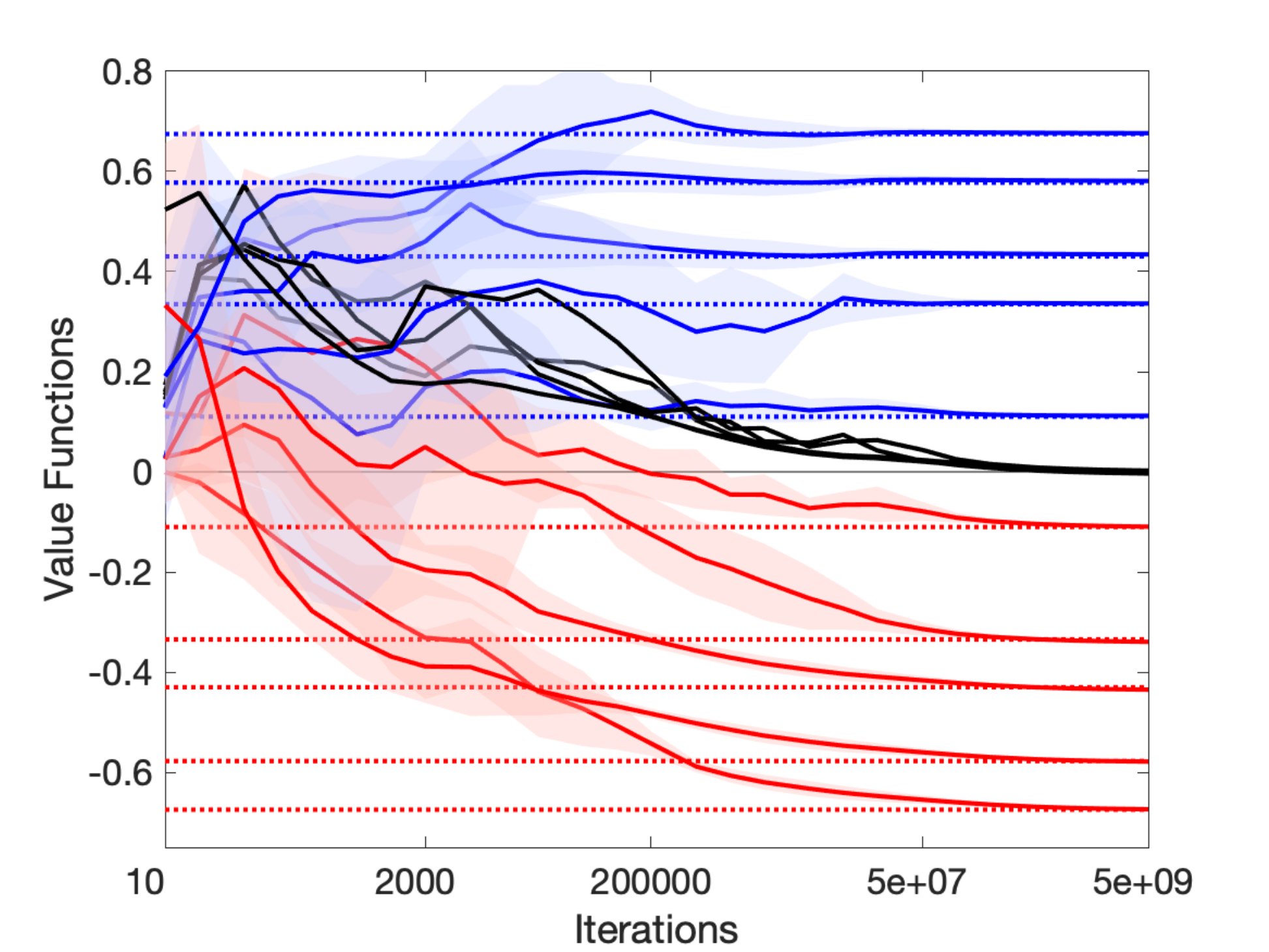}
		  \centering{\small (b)  Convergence under Assumption \ref{assume:second}}
	\end{minipage}   
	\caption{Convergence of the value function estimates after $20$ runs of the decentralized $Q$-learning dynamics. The \textbf{\textcolor{red}{red}}  and  \textbf{\textcolor{blue}{blue}} curves represent the quantities for players $1$ and $2$, respectively. The solid \textbf{\textcolor{red}{red}}/\textbf{\textcolor{blue}{blue}} lines and the bar-areas around them denote the average value function estimates $\{\hat{v}_{s,k}^i\}_{k\geq 0}$  at all  states $s\in S$ and the standard deviations, after the $20$ runs, respectively. The solid \textbf{black} lines denote the summation of the estimates $\{\hat{v}_{s,k}^1+\hat{v}_{s,k}^2\}_{k\geq 0}$ at each $s$ (which should converge to (almost) zero asymptotically). The dotted lines  {\color{red}\dotted}/{\color{blue}\dotted} denote the actual Nash equilibrium values; the dashed-dotted lines {\color{red}\chain}/{\color{blue}\chain} denote the boundaries of the neighborhoods around the Nash equilibrium given in Theorem \ref{thm:main}. 
	(a)  Convergence to the neighborhood of the Nash equilibrium value,  under Assumption \ref{assume:first}. (b) Convergence to the Nash equilibrium value, under Assumption \ref{assume:second}.}
\label{fig:simulation_results}
\vspace{-5pt}
\end{figure*}

\begin{figure*}[!t]
	\centering
	\hspace{-20pt}
	\begin{minipage}[c]{0.5\linewidth}
		\includegraphics[width=\linewidth]{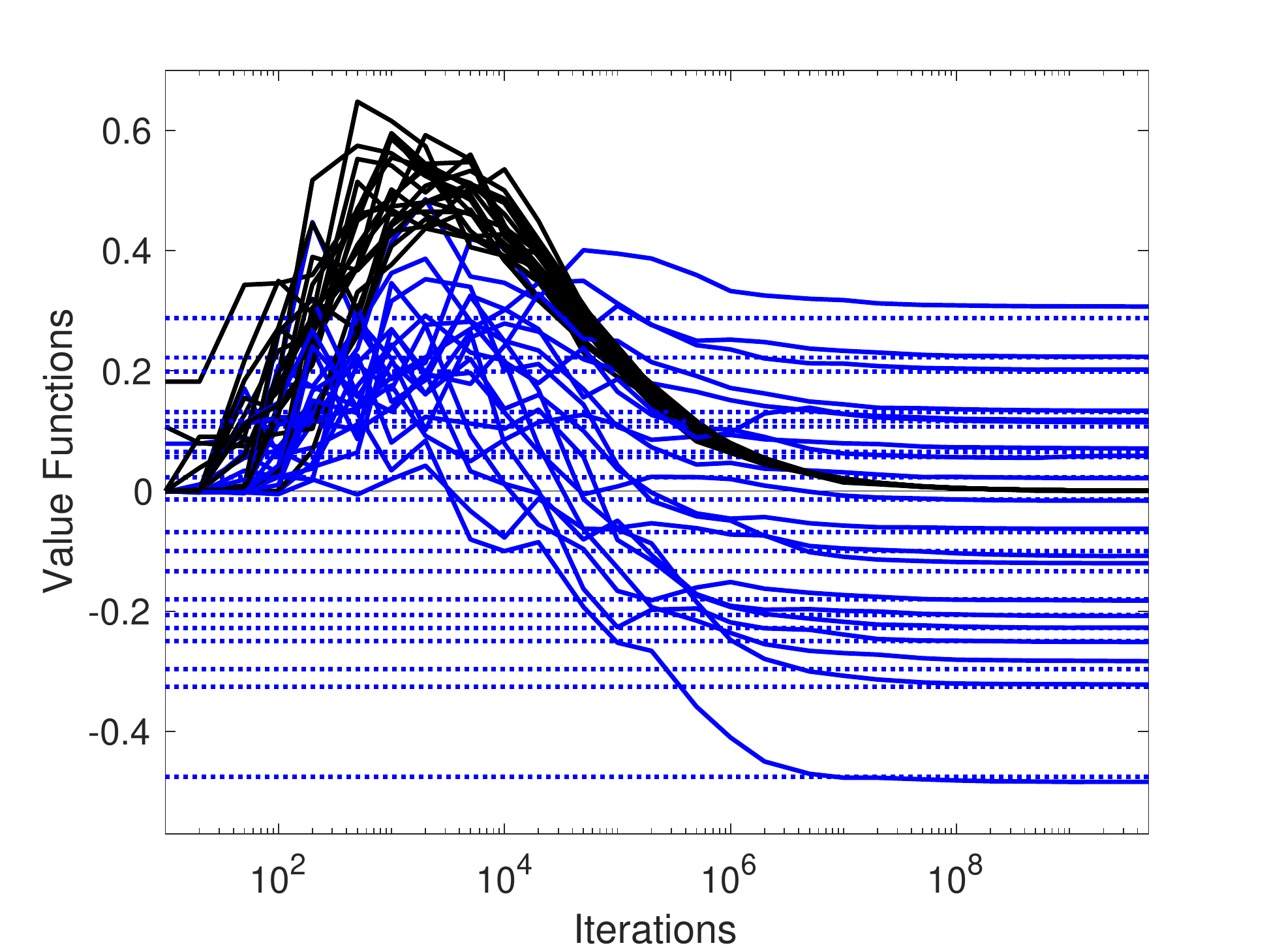}
		 \centering{\small(a) Convergence under Assumption \ref{assume:first}}
	\end{minipage} 
	\hspace{-19pt}
	\begin{minipage}[c]{0.5\linewidth}
		\includegraphics[width=\linewidth]{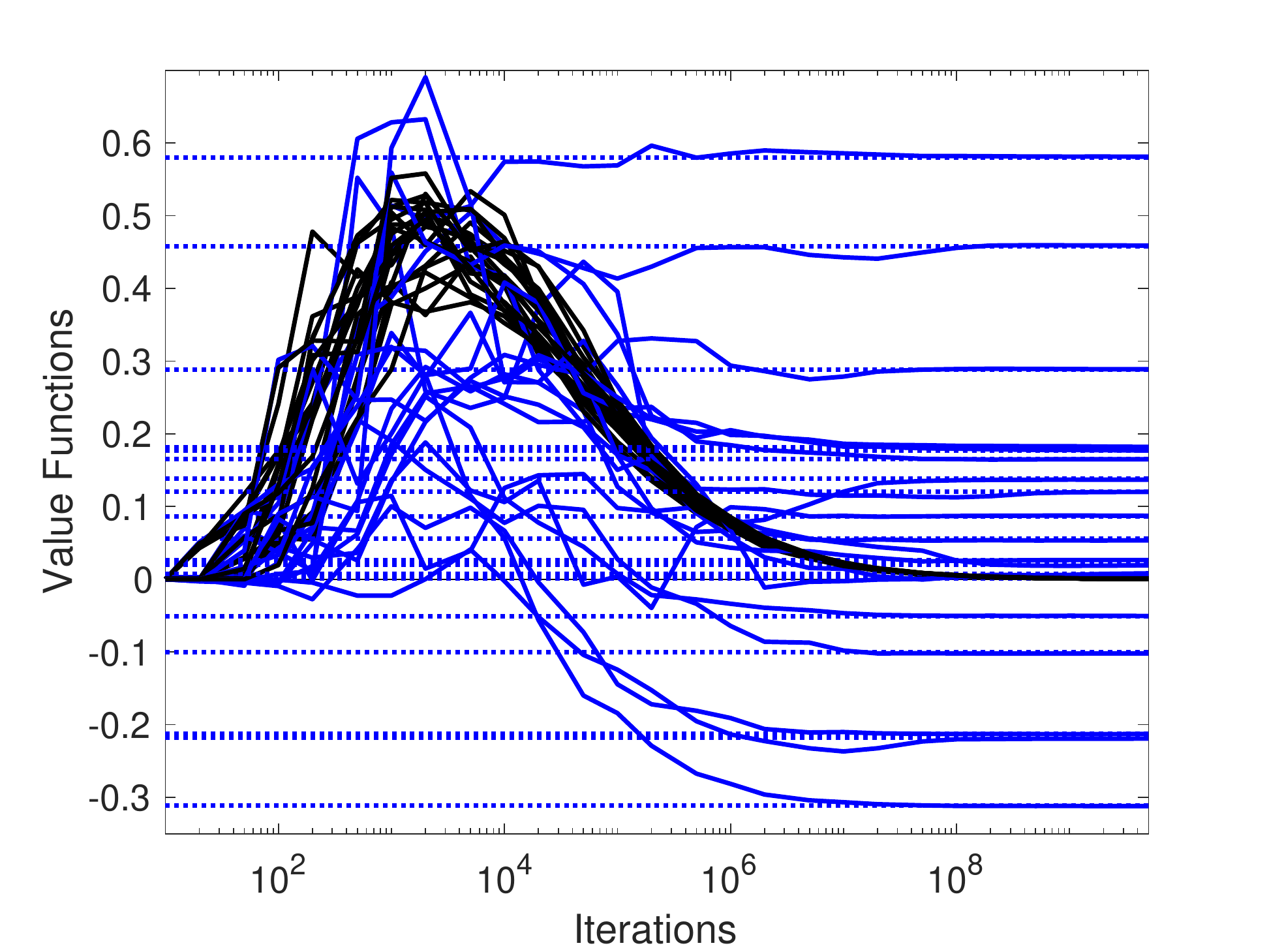}
		  \centering{\small (b)  Convergence under Assumption \ref{assume:second}}
	\end{minipage}   
	\caption{Convergence of the value function estimates of the decentralized $Q$-learning dynamics.  The  \textbf{\textcolor{blue}{blue}} curves represents the quantities for player $2$.  The solid \textbf{\textcolor{blue}{blue}} lines  denote the  value function estimates $\{\hat{v}_{s,k}^2\}_{k\geq 0}$  at all  states $s\in S$. The solid \textbf{black} lines denote the summation of the estimates $\{\hat{v}_{s,k}^1+\hat{v}_{s,k}^2\}_{k\geq 0}$ at each $s$ (which should converge to (almost) zero asymptotically). The dotted lines {\color{blue}\dotted} denote the actual Nash equilibrium values. 
	(a)  Convergence to the neighborhood of the Nash equilibrium value,  under Assumption \ref{assume:first}. (b) Convergence to the Nash equilibrium value, under Assumption \ref{assume:second}.}
\label{fig:simulation_results_2}
\vspace{-5pt}
\end{figure*}

 All the simulations are executed on a desktop computer equipped with a 3.7 GHz Hexa-Core Intel Core i7-8700K processor with Matlab R2019b. The device also has two 8GB 3000MHz DDR4 memories and a NVIDIA GeForce GTX 1080 8GB GDDR5X graphic card. For illustration, we consider a zero-sum Markov game with $5$ states and $3$ actions at each state, i.e., $S=\{1,2,\cdots,5\}$ and $A_s^i=\{1,2,3\}$. The discount factor  $\gamma=0.6$. The reward functions are chosen randomly in a way that $r^1_s(a^1,a^2)\propto \bar{r}_{s,a^1,a^2}\cdot\exp{(s^2)}$ for $s\in S$, where $\bar{r}_{s,a^1,a^2}$ is uniformly drawn from $[-1,1]$. $r^1_s(a^1,a^2)$ is then normalized by $\max_{s,a^1,a^2}\{r^1_s(a^1,a^2)\}$ so that $|r_s^i(a^1,a^2)|\leq R=1$ for all $(i,s,a^1,a^2)$. For the state transition dynamics $p$, we construct two cases, \textbf{Case 1} and \textbf{Case 2} by randomly generating transition probabilities, so that  they  satisfy Assumptions  \ref{assume:firstIO} and \ref{assume:secondIO}, respectively. For both cases, we choose $\alpha_c=1/c^{0.9}$ and $\beta_c=1/c$ with $\rho_\alpha=0.9$, $\rho_\beta=1$, and $\rho=0.7$, and set $\tau_c$ in accordance with  \eqref{equ:tau_to_epsilon} and \eqref{equ:tau_to_zero}, respectively. For \textbf{Case 1}, we choose $\epsilon=2\times 10^{-4}$ and $\bar{\tau}=4.5\times 10^4$; for \textbf{Case 2}, we choose $\bar{\tau}=0.07$. Note that the different  choices of $\bar{\tau}$ are due to the different decreasing rates of $\tau_c$ in \eqref{equ:tau_to_epsilon} and \eqref{equ:tau_to_zero}, and are chosen to generate aesthetic plots.  The simulation results are illustrated in Figure \ref{fig:simulation_results}, which are generated after $20$ runs of  our learning dynamics. 

As shown in Figure \ref{fig:simulation_results} (a), for \textbf{Case 1}, the value function estimates successfully converge to the neighborhood of the Nash equilibrium values, where the size of the neighborhood is indeed controlled by \eqref{equ:result_to_eps}. For  \textbf{Case 2}, it is shown in Figure \ref{fig:simulation_results} (b)  that the value function estimates converge to the Nash equilibrium, as $\tau\to 0$.  These observations have corroborated our theory established in \S\ref{sec:main}. Moreover, it is observed that the variance of the iterates decreases as they converge to the (neighborhood of) Nash equilibrium, implying the almost-sure convergence guarantees we have  established. 

Besides the illustrative example, we have also tested our learning dynamics on larger scale games, and validated our theory for this  case (see Figure \ref{fig:simulation_results_2}).  See \S\ref{sec:add_sim} for more details of the example. 


\section{Concluding Remarks} \label{sec:conclusion}

This paper has studied decentralized multi-agent reinforcement learning in zero-sum Markov games. We have developed a decentralized $Q$-learning dynamics with provable convergence guarantees to the (neighborhoods of the) Nash equilibrium value of the game. Unlike many existing MARL algorithms, our learning dynamics is both rational and convergent, and only based on the payoffs received and the local actions executed by each agent. It also requires neither asymmetric stepsizes/update rules or any coordination for the agents, nor even being aware of the existence of the opponent. 
 
\begin{ack}
M. O. Sayin was with the Laboratory for Information and Decision Systems at MIT when this paper was submitted. K.  Zhang and A. Ozdaglar were  supported by DSTA grant  031017-00016. T. Ba\c{s}ar was supported in part by ONR MURI Grant N00014-16-1-2710 and in part by AFOSR Grant FA9550-19-1-0353. 
\end{ack}

\bibliographystyle{plainnat}
\bibliography{myref}

\clearpage 

\appendix

\onecolumn

~\\
\centerline{{\fontsize{13.5}{13.5}\selectfont \textbf{Supplementary Materials for ``Decentralized  $Q$-Learning}}}

\vspace{6pt}
 \centerline{\fontsize{13.5}{13.5}\selectfont \textbf{
  in Zero-sum Markov Games''}}
\vspace{6pt}


\section{Related Work}\label{sec:relatedwork} 

We here focus on the related works on MARL with provable convergence guarantees. 
 
\subsection{Multi-agent RL in Markov Games}

Stemming from the seminal work \cite{littman1994markov}, Markov games have been widely recognized as the benchmark setting  for MARL. \cite{littman1994markov} focused on the zero-sum setting, and developed minimax $Q$-learning algorithm with asymptotic  converge guarantees  \citep{szepesvari1999unified}. However, this algorithm requires each agent to observe the opponent's action. More importantly, each agent is fully aware of the zero-sum game being played, and solves a linear program to solve a matrix game  at each iteration. 
Subsequently, \cite{bowling2001rational} proposed that a preferable MARL algorithm should be both \emph{rational} and \emph{convergent}: a rational algorithm ensures that the iterates converge to the  opponent's best-response if the opponent converges to a stationary policy; while a convergent algorithm ensures convergence to some equilibrium  if all the agents apply the learning dynamics. In this sense,  minimax $Q$-learning is not rational. In contrast, our learning dynamics is both rational and convergent. 

In the same vein as minimax $Q$-learning, with coordination among agents, asymptotic convergence has also been established for other $Q$-learning variants beyond the zero-sum setting \citep{littman2001friend,hu2003nash,greenwald2003correlated}. \cite{ref:Borkar02} has also established the asymptotic  convergence of an actor-critic algorithm to a weaker notion of generalized Nash equilibrium. 
Recently, there is an increasing interest in studying the \emph{non-asymptotic}  performance of MARL in Markov games 
\citep{perolat2015approximate,wei2017online,sidford2019solving,xie2020learning,bai2020provable,bai2020near,zhang2019policy,zhang2020model,shah2020reinforcement,liu2020sharp,zhao2021provably}. These algorithms are in essence centralized, in that they require either  the control of both agents \citep{perolat2015approximate,sidford2019solving,xie2020learning,bai2020provable,bai2020near,zhang2019policy,zhang2020model,shah2020reinforcement,liu2020sharp,zhao2021provably}, or at least the observation of the opponent's actions  \citep{wei2017online,xie2020learning}. 
 
Two closely related recent papers \cite{ref:Leslie20} and \cite{ref:Sayin20} have presented, respectively, continuous-time best response dynamics and discrete-time fictitious play dynamics that can converge to an equilibrium in zero-sum Markov games. They have established provable convergence by addressing the non-stationarity issue through a two-timescale framework. Though these two-timescale dynamics share a similar flavor with our approach, still, observing the opponent's mixed strategy (in \cite{ref:Leslie20}) or actions (in \cite{ref:Sayin20}) is indispensable in them and plays an important role in their analysis. This is in stark contrast to our dynamics that require minimal information, i.e., being radically uncoupled \citep{ref:Foster06,ref:Leslie05}.  

\subsection{Decentralized Multi-agent Learning} \label{sec:dec_MARL_related_work}

Decentralized learning is a desired property, and has been studied for matrix games (single-state Markov games) under the framework of \emph{no-regret learning}  \citep{cesa2006prediction,freund1999adaptive,mertikopoulos2019learning}. \cite{ref:Leslie05} also proposed individual soft $Q$-learning dynamics  for zero-sum matrix games. For general Markov games, however, it is  known that blindly  applying  independent/decentralized $Q$-learning can easily diverge, due to the \emph{non-stationarity} of the environment   \citep{tan1993multi,boutilier1996planning,matignon2012independent}. 
Despite this, the decentralized paradigm has still attracted continuing research interest  \citep{ref:Arslan17,perolat2018actor,daskalakis2020independent,tian2020provably,wei2021last}, since it is much more scalable and natural for agents to implement. Notably, these works are not as decentralized and as general as  our learning dynamics.    
 
Specifically, the algorithm in \cite{ref:Arslan17} requires the agents to coordinately explore every multiple iterations (the exploration phase),  without changing their policies within each exploration phase, in order to create a stationary environment for each agent. Similar to our work, \cite{perolat2018actor} also proposed decentralized and two-timescale algorithms, which, however, is an actor-critic algorithm where the value functions are estimated at a  faster timescale (critic step), and the policy is improved at a slower one (actor step). More importantly, the algorithm only applies to Markov games with a ``multistage'' structure, in which each state can only be visited once. Establishing convergence in general zero-sum Markov games is posted as an open problem in \cite{perolat2018actor}.  
In \cite{daskalakis2020independent}, the agents have to coordinate to  use two-timescale stepsizes in the updates. In contrast, our learning dynamics does not require any coordination among agents, and each agent plays a symmetric role in learning, referred to as \emph{strongly independent} in  \cite{daskalakis2020independent}. In fact, developing provable guarantees for strongly independent algorithms is considered as an important open question in \cite{daskalakis2020independent}.

Two recent works \cite{tian2020provably,wei2021last} studied the decentralized setting that is closest to ours. 
\cite{tian2020provably} focused on the exploration aspect for finite-horizon settings, and considered a weak notion of regret. It is unclear if the learning dynamics converge to any equilibrium when both agents apply it\footnote{Note that the same update rule with different stepsize and bonus choices and a  certified policy technique, however, can return a non-Markovian approximate Nash equilibrium policy pair in  the self-play setting, by storing the whole history of the learning process; see \cite{bai2020near} for more details.}. Contemporaneously,  \cite{wei2021last} presented an interesting optimistic variant of the  gradient descent-ascent method, with  a strong guarantee of last-iterate convergence rates,  which shares all the desired properties as our learning dynamics. The algorithm is  delicately designed and different from the common value/policy-based RL update rules, e.g., $Q$-learning, as in our work. Moreover, to characterize finite-time convergence, in the model-free setting, the agents need to coordinate to interact multiple steps at each iteration of the algorithm, while our learning dynamics is coordination-free with natural update rules. These two works can thus  be viewed as orthogonal to ours. 

After submitting the first draft of our paper, we were reminded of an independent and concurrent work of \cite{guo2021decentralized}, which also studied a decentralized learning setting in zero-sum Markov games. We summarize the substantial differences between the two works as follows. 

\textbf{Motivation:} In \cite{guo2021decentralized}, being ``decentralized'' is defined as ``each player not knowing the opponent's action'', to ``protect the privacy'', and the goal is to ``compute'' the Nash equilibrium of the game; in contrast, in our work, in addition to ``being oblivious to the opponent's action'', we also allow no ``coordination'' among agents, so that each agent can simply run the learning dynamics ``individually'', without even being aware of the existence of the opponent. The agents in our setting are considered as self-interested decision-makers, who seek to adapt to the opponent's play by inferring it from the rewards received without seeing the opponent's actions. The Nash equilibrium, on the other hand, is the result that ``emerge'' naturally when both agents follow this self-interested learning dynamics (and we have proved this). Finally, as our learning dynamics are oblivious to the opponent and are adaptive to the opponent, we expect it to converge beyond the zero-sum setting (e.g., the identical-interest setting), which is one of our ongoing research directions. In contrast, the algorithm in \cite{guo2021decentralized} is specifically developed for the zero-sum setting. 
These motivations differ fundamentally from \cite{guo2021decentralized}  (and thus creates very different technical challenges, as detailed below).
 
\textbf{Learning dynamics (Algorithms):} The algorithm in \cite{guo2021decentralized},  is actor-critic, which is a type of policy-based RL method; the learning dynamics in our work is Q-learning based, which belongs to value-based RL methods. More importantly, the update-rule in \cite{guo2021decentralized},  is of ``double-loop'' form, in the sense that it fixes the iterate of Player 1 while updating Player 2's policy, so that a ``best-response'' policy of Player 2 can be obtained. This is an asymmetric update-rule, and requires coordination between agents. In contrast, our learning dynamics are ``symmetric'', without such a double-loop coordination, where each agent simply runs her own $Q$-learning dynamics. 

\textbf{Assumptions and results:}  \cite{guo2021decentralized}  considers a function approximation setting, and assumes that: 1) the ``double-loop'' update can be implemented by the agents in the decentralized setting; 2) the concentration (or  ``Concentrability'') coefficient is finite (Assumption 4.1), for ``an arbitrary sequence of policies''; 3) samples are drawn i.i.d. from the stationary state-action distribution; 4) projection of the iterates onto some ball with radius $R$, to ensure the iterates' stability; and 5) zero approximation error of the Bellman operator (Assumption 4.2). Under these assumptions, non-asymptotic convergence results were established. In contrast, our work considers a fundamental tabular setting, and without making these assumptions (1-4), with instead asymptotic convergence guarantees.  
 With these significantly different assumptions, it is not clear if one paper's  result implies  the other's. 
 
\textbf{Analysis techniques (Technical novelty):} The analyses, as well as the technical novelties in both papers are not comparable. The analysis technique in \cite{guo2021decentralized},  is a mirror-descent type of analysis, based on the convergence analysis of policy gradient (and actor-critic) algorithms in single-agent RL. 
The techniques in our paper, however, are based on stochastic approximation theory, a classic technique in showing the convergence of $Q$-learning. The challenges we need to address (our technical novelties) mainly lie in constructing a Lyapunov function and stability of the iterates, within this non-standard two-timescale stochastic approximation setting, with asynchronous updates. Such challenges would not be encountered in the analysis of \cite{guo2021decentralized}, making the technical novelties of the two papers fundamentally different. 


\section{Examples}\label{sec:example}

In this section, we provide three sets of parameter examples and highlight whether they satisfy Assumptions \ref{assume:common} and \ref{assume:first2} or Assumptions \ref{assume:common} and \ref{assume:second2}. Recall that Assumptions \ref{assume:firstIO} and \ref{assume:secondIO} do not impose conditions on the step sizes nor the temperature parameter.

\begin{example}\label{ex:1}
Set the step sizes as $\alpha_c=c^{-\rhoa}$, $\beta_c=c^{-\rhob}$, where $1/2<\rhoa<\rhob\leq 1$ and the temperature parameter as
\begin{equation}
\tau_c = \frac{1}{c}\bar{\tau} + \left(1-\frac{1}{c}\right)\epsilon
\end{equation}
for some $\bar{\tau}>0$.
\end{example}

In the following, we show that Example \ref{ex:1} satisfies Assumption \ref{assume:common} and \ref{assume:first2}. Assumption \ref{assume:common1} holds since $\sum_{c>0}(1/c)^{\rho}$ is convergent if $\rho>1$, and divergent if $\rho\leq1$. Assumption \ref{assume:common2} holds since there exists a non-decreasing polynomial, e.g., $C(x)= M_ox^{m}$, for all $x\geq 1$, where $M_o:=M^{-\rhob/(\rhob-\rhoa)}>0$ and $m\in\{m'\in\mathbb{Z}_+: m'\geq 1/(\rhob-\rhoa)\}$. Particularly, we have
\begin{equation}
\frac{\beta_l}{\alpha_c}=\frac{c^{\rhoa}}{l^{\rhob}} > \lambda \quad \Leftrightarrow \quad l < \left(\frac{1}{\lambda}\right)^{\frac{1}{\rhob}}c^{\frac{\rhoa}{\rhob}}.
\end{equation}
We claim that $M c\geq \lambda^{-1/\rhob}c^{\rhoa/\rhob}$ for all $c\geq C(\lambda^{-1})$ because $\lambda^{-1/\rhob}c^{\rhoa/\rhob} \leq M c$ yields
\begin{equation}
 c^{1-\frac{\rhoa}{\rhob}}\geq \left(\frac{1}{\lambda}\right)^{\frac{1}{\rhob}} \frac{1}{M} \quad \Leftrightarrow \quad c\geq \left(\frac{1}{\lambda}\right)^{\frac{1}{\rhob-\rhoa}} \left(\frac{1}{M}\right)^{\frac{\rhob}{\rhob-\rhoa}}= M_o \left(\frac{1}{\lambda}\right)^{\frac{1}{\rhob-\rhoa}}.
\end{equation}
On the other hand, we have $C(\lambda^{-1}) = M_o \lambda^{-m} \geq M_o \lambda^{-1/(\rhob-\rhoa)}$ for all $\lambda \in(0,1)$ since $m\geq 1/(\rhob-\rhoa)$ by its definition. 

Assumption \ref{assume:first2} holds since $\{\tau_c\}$ monotonically decreases to $\epsilon$ as $c\rightarrow\infty$, and 
\begin{align}
\frac{\tau_{c+1}-\tau_c}{\alpha_c} &= c^{\rhoa}\left[\frac{1}{c+1}\bar{\tau} + \left(1-\frac{1}{c+1}\right)\epsilon - \frac{1}{c}\bar{\tau} - \left(1-\frac{1}{c}\right)\epsilon\right]\\
&=\frac{c^{\rhoa}}{c(c+1)}(\epsilon-\bar{\tau}),
\end{align}
which goes to zero as $c\rightarrow\infty$ since $\rhoa<1$, and $\sum_{c>0}c^{-2\rhoa} < \infty$ since $2\rhoa>1$. \hfill $\square$

\begin{example}\label{ex:2}
Set the step sizes as $\alpha_c=c^{-\rhoa}$, $\beta_c=c^{-\rhob}$, where $1/2<\rhoa<\rhob\leq 1$ and the temperature parameter as
\begin{equation}\label{eq:tauc}
\tau_c' = \bar{\tau}\left(1+\bar{\tau}\frac{\rhoa\rho}{4D} \log(c)\right)^{-1}
\end{equation}
for some $\rho\in(0,2-1/\rhoa)$ and $\bar{\tau}>0$.
\end{example}

In the following, we show that Example \ref{ex:2} satisfies Assumption \ref{assume:common} and \ref{assume:second2}. Example \ref{ex:2} shares the same step sizes with Example \ref{ex:1}. Therefore, Assumption \ref{assume:common} holds as shown above for Example \ref{ex:1}. On the other hand, Assumption \ref{assume:second2} also holds since $\{\tau_c'\}$ monotonically decreases to $0$ as $c\rightarrow\infty$, and
\begin{equation}\label{eq:tuu}
0\geq \frac{\tau_{c+1}'-\tau_c'}{\alpha_c} \geq \frac{c^{\rhoa}(\log(c+1)-
\log(c))}{\log(c+1)\log(c)}, 
\end{equation}
where the right-hand side goes to zero as $c\rightarrow\infty$, and
\begin{align}
\alpha_c^{\rho}\exp\left(\frac{4D}{\tau_c'}\right)&=c^{-\rhoa\rho}\exp\left(\frac{4D}{\bar{\tau}}\left(1+\bar{\tau}\frac{\rhoa\rho}{4D}\log(c)\right)\right)\\
&=c^{-\rhoa\rho}\exp\left(\frac{4D}{\bar{\tau}}\right)\exp\left(\rhoa\rho\log(c)\right)\\
&=\exp\left(\frac{4D}{\bar{\tau}}\right),\quad \forall c>0,
\end{align}
which implies that $\alpha_c^{\rho}\exp(4D/\tau_c')\leq C'$ for all $c>0$ when $C' = \exp(4D/\bar{\tau})$. \hfill $\square$

\begin{example}\label{ex:3}
Set the step sizes as $\alpha_c=c^{-\rhoa}$, $\beta_c=c^{-\rhob}$, where $1/2<\rhoa<\rhob\leq 1$ and the temperature parameter as $\tau_c = \max\{\epsilon,\tau_c'\}$, where $\tau_c'$ is as described in \eqref{eq:tauc}.
\end{example}

In the following, we show that Example \ref{ex:1} satisfies Assumption \ref{assume:common} and \ref{assume:first2}. Example \ref{ex:3} shares the same step sizes with Example \ref{ex:1}. Therefore, Assumption \ref{assume:common} holds as shown above for Example \ref{ex:1}. On the other hand, Assumption \ref{assume:first2} also holds since $\{\tau_c\}$ monotonically decreases to $\epsilon$ as $c\rightarrow\infty$ (which follows since $\{\tau_c'\}$ monotonically decreases to $0$ as $c\rightarrow\infty$), and we again have the inequality \eqref{eq:tuu} and $\sum_{c>0}c^{-2\rhoa} < \infty$ since $2\rhoa>1$. \hfill $\square$

\section{Proofs of Propositions \ref{prop:bound}-\ref{prop:io}}\label{sec:props}

\begin{lemma*}{Proposition \ref{prop:bound}}
Since $\bar{\alpha}_{k}^i\in(0,1]$ for all $k\geq 0$, $\beta_c\in(0,1)$ for all $c>0$, $\|\hat{q}_{s,0}^i\|_{\infty}\leq D$ and $|\hat{v}_{s,0}^i|\leq D$ for all $(i,s)$, the iterates are bounded, e.g., $\|\hat{q}_{s,k}^i\|_{\infty}\leq D$ and $|\hat{v}_{s,k}^i|\leq D$ for all $(i,s)$ and $k\geq 0$.
\end{lemma*}

\begin{myproof}
The proof follows from the fact that the initial iterates are picked within the compact set and they continue to remain inside it since they are always updated to a convex combination of two points inside.
\end{myproof}

\begin{lemma*}{Proposition \ref{prop:reduce}}
Suppose that Assumption \ref{assume:common1} and either Assumption \ref{assume:first2} or \ref{assume:second2} hold. Then, there exists $C_s\in\mathbb{Z}_+$ for each $s\in S$ such that $\alpha_c\exp\left(2D/\tau_c\right)< \min_{i=1,2}\{|A_s^i|^{-1}\}$, for all $c\geq C_s$. Correspondingly, the update of the local $Q$-function estimate \eqref{eq:q} reduces to
\begin{equation*}
\hat{q}_{s_k,k+1}^i[a_{k}^i] = \hat{q}_{s_k,k}^i[a_{k}^i] + \frac{\alpha_{\#s_k}}{\smooth_{k}^i[a_{k}^i]}\left(r_{k}^i + \gamma \hat{v}_{s_{k+1},k}^i - \hat{q}_{s_k,k}^i[a_{k}^i]\right),
\end{equation*}
for all $\#s_k \geq C_{s_k}$ since $\alpha_{\#s_k}/\smooth_{k}^i[a_{k}^i]\leq |A_{s_k}^i|\alpha_{\#s_k}\exp\left(2D/\tau_{\#s_k}\right)\leq 1$ by \eqref{eq:def_smooth_br_2}.
\end{lemma*}

\begin{myproof}
If Assumption \ref{assume:first2} holds, then $\alpha_c\exp(2D/\tau_c)\leq \alpha_c \exp(2D/\epsilon)$. Since $\alpha_c\rightarrow 0$ as $c\rightarrow\infty$ by Assumption \ref{assume:common1}, there exists such $C_s$.

If Assumption \ref{assume:second2} holds, then 
\begin{align}
\alpha_c\exp\left(\frac{2D}{\tau_c}\right)&=\alpha_c^{1-\rho/2}\left(\alpha^{\rho}\exp\left(\frac{4D}{\tau_c}\right)\right)^{1/2}\\
&\leq \alpha_c^{1-\rho/2}\sqrt{C'},\quad \forall c\geq C.
\end{align}
Since $1-\rho/2 >0$ and $\alpha_c\rightarrow 0$ as $c\rightarrow\infty$, there exists such $C_s$.
\end{myproof}

\begin{lemma*}{Proposition \ref{prop:io}}
Suppose that either Assumption \ref{assume:first} or Assumption \ref{assume:second} holds. Then, at any stage $k$, there is a fixed positive probability, e.g., $\underline{p}>0$, that the game visits any state $s$ at least once within $n$-stages independent of how players play. Therefore, $\#s\rightarrow \infty$ as $k\rightarrow\infty$ with probability $1$. 
\end{lemma*}

\begin{myproof}
By Borel-Cantelli Lemma, if we have
\begin{equation}
\sum_{k\geq 0} \Prob{\#_ks\leq \lambda}<\infty,\quad\forall \lambda\in\mathbb{N},\label{eq:Borel1}
\end{equation}
then we have $\#_ks\rightarrow\infty$ as $k\rightarrow\infty$ ith probability $1$. To show \eqref{eq:Borel1}, we partition the time axis into $n$-stage intervals and introduce an auxiliary counting process $\bar{\#}_k^ns$ that increases by $1$ at the end of each interval if state $s$ is visited at least once within the last $n$-stages. By its definition, we have $\bar{\#}_k^ns\leq \#_k s$. Correspondingly, we have $\Prob{\#_ks\leq \lambda}\leq \Prob{\bar{\#}_k^ns\leq \lambda}$. The right-hand side is one for all $k<n\lambda $. On the other hand, for $k\geq n\lambda$, we have
\begin{align}
\Prob{\bar{\#}_k^ns\leq \lambda} &\leq \sum_{l=0}^{\lambda}\binom{\left\lfloor \frac{k}{n}\right\rfloor}{l} 1^l(1-\underline{p})^{\left\lfloor\frac{k}{n}\right\rfloor-l}\\
&\leq (1-\underline{p})^{\left\lfloor\frac{k}{n}\right\rfloor-\lambda}\sum_{l=0}^{\lambda}\binom{\left\lfloor \frac{k}{n}\right\rfloor}{l}\label{eq:p3b1}
\end{align}
since $(1-\underline{p})<1$. 

Next, we can resort to the following inequality \cite[Lemma 16.19]{ref:Flum06book}
\begin{equation}\label{eq:p3b2}
\sum_{l=0}^{L} \binom{k}{l}\leq 2^{H_2\left(L/k\right)k},\quad\mbox{if } L/k\leq 1/2,
\end{equation} 
where $H_2(p):=-p\log(p)-(1-p)\log(1-p)$. Therefore, for $k\geq 2n\lambda$, \eqref{eq:p3b1} and \eqref{eq:p3b2} yield that
\begin{align}
\Prob{\bar{\#}_k^ns\leq \lambda} &\leq \frac{1}{(1-\underline{p})^{\lambda}}\left[(1-\underline{p})2^{H_2(\lambda/\lfloor k/n\rfloor)}\right]^{\lfloor k/n\rfloor}.
\end{align}
Since $H_2(p)\geq 0$ is an increasing continous function for $p\in(0,0.5)$ and $H_2(0)=0$, there exists $\kappa\in\mathbb{N}$ such that $(1-\underline{p})2^{H_2(\lambda/\lfloor k/n\rfloor)}< (1-\underline{p})2^{H_2(\lambda/\lfloor \kappa/n\rfloor)}<1$ for all $k\geq \kappa$. Define $\xi:= (1-\underline{p})2^{H_2(\lambda/\lfloor \kappa/n\rfloor)}$. Then we have
\begin{align}
\sum_{k\geq 0} \Prob{\#_ks\leq \lambda} \leq \kappa + \frac{1}{(1-\underline{p})^{\lambda}}\sum_{k=\kappa+1}^{\infty}\xi^{\lfloor k/n\rfloor}\leq \kappa + \frac{1}{\xi(1-\underline{p})^{\lambda}}\sum_{k=\kappa+1}^{\infty}(\xi^{1/n})^{k}
\end{align}
and the right-hand side is convergent since $\xi^{1/n} <1$, which completes the proof.
\end{myproof}

\section{Preliminary Information on Stochastic Approximation Theory}\label{sec:prelim}

Here, we present two preliminary results. The former uses a continuous-time approximation to analyze a discrete-time update \citep{ref:Benaim99}. The latter is about characterizing the convergence properties of an asynchronous discrete-time update by exploiting certain bounds on their evolution \citep{ref:Sayin20}.

\subsection{Stochastic Approximation via Lyapunov Function}\label{pre:Lyapunov}

The following theorem (follows from \cite[Proposition 4.1 and Corollary 6.6]{ref:Benaim99}) characterizes the conditions sufficient to characterize the convergence properties of a discrete-time update:
\begin{equation}\label{eq:discrete}
x_{k+1} = x_k + \lambda_k \left[F(x_k) + \epsilon_k + \omega_{k}\right],
\end{equation}
through its limiting ordinary differential equation (o.d.e.):
\begin{equation}\label{eq:ode}
\frac{dx(t)}{dt} = F(x(t)).
\end{equation}

\begin{theorem}\label{thm:Lyapunov}
Suppose that there exists a Lyapunov function $V:\mathbb{R}^m\rightarrow [0,\infty)$ for \eqref{eq:ode}.\footnote{We say that a continous function $V$ is a {\em Lyapunov function} for a flow provided that for any trajectory of the flow, e.g., $x(t)$, $V(x(t'))<V(x(t))$ for all $t'>t$ if $V(x(t))>0$ else $V(x(t'))=0$ for all $t'>t$.} Furthermore, 
\begin{itemize}
\item[$i)$] The step sizes $\{\lambda_k\in [0,1]\}_{k=0}^{\infty}$ decrease at a suitable rate:
\begin{equation}
\sum_{k=0}^{\infty} \lambda_k = \infty \mbox{ and }\sum_{k=0}^{\infty}\lambda_k^2 < \infty.
\end{equation} 
\item[$ii)$] The iterates $x_k\in\mathbb{R}^m$, for $k=0,1,\ldots$, are bounded, e.g., $\sup_k \|x_k\|_{\infty} < \infty$.
\item[$iii)$] The vector field $F:\mathbb{R}^m\rightarrow \mathbb{R}^m$ is globally Lipschitz continous.
\item[$iv)$] The stochastic approximation term $\omega_{k}\in\mathbb{R}^m$ satisfies the following condition for all $K>0$,\footnote{This is a more general condition than assuming that $\{\omega_{k}\}$ is a square-integrable Martingale difference sequence, e.g., see \cite[Section 2]{ref:Borkar08}.}
\begin{equation}\label{eq:omegacond}
\lim_{k\rightarrow \infty} \sup_{n> k : \sum_{l=k}^{n-1}\lambda_l \leq K}\left\{\left\|\sum_{l=k}^{n-1}\lambda_l\omega_{l}\right\|\right\} = 0.
\end{equation}
\item[$v)$] The error term $\epsilon_k\in\mathbb{R}^m$ is asymptotically negligible, i.e., $\lim_{k\rightarrow\infty}\|\epsilon_k\| = 0$, with probability $1$.
\end{itemize}
Then the limit set of \eqref{eq:discrete} is contained in the set
\begin{equation}
\{x\in\mathbb{R}^m:V(x)=0\},
\end{equation}
with probability $1$.
\end{theorem}

\subsection{Asynchronous Stochastic Approximation} 

Consider the scenarios where we update only a subset of entries of the iterate $x_k$ with specific step sizes. For example, the $l$th entry of the iterate $x_k\in\mathbb{R}^m$, denoted by $x_k[l]$, gets updated only at certain (possibly random) time instances with a specific step size $\lambda_{k,l}\in[0,1]$. Furthermore, there is not necessarily a time-invariant vector field $F$ as in Theorem \ref{thm:Lyapunov}. The following theorem, \cite[Theorem 3]{ref:Sayin20}, characterizes the limit set of $\{x_k\}_{k\geq 0}$ provided that it evolves within a shrinking envelope.\footnote{This theorem is a rather straight-forward modification of \cite[Theorem 3]{ref:Tsitsiklis94}.}

\begin{theorem}\label{thm:bounds}
Suppose that the evolution of $\{x_k\}_{k\geq 0}$ always satisfies the following upper and lower bounds:
\begin{subequations}\label{eq:bounds}
\begin{align}
&x_{k+1}[l] \leq (1-\lambda_{k,l})x_{k}[l] + \lambda_{k,l} (\gamma \|x_k\|_{\infty} + \overline{\epsilon}_{k}),\\
&x_{k+1}[l] \geq (1-\lambda_{k,l})x_{k}[l] + \lambda_{k,l} (-\gamma \|x_k\|_{\infty} + \underline{\epsilon}_{k}),
\end{align}
\end{subequations}
where $\gamma\in(0,1)$ is a discount factor, $\|x_k\|_{\infty}\leq D$ for all $k\geq 0$ for a fixed $D$, and the specific step sizes satisfy the usual conditions: 
\[\sum_{k=0}^{\infty} \lambda_{l,k} = \infty\quad \mbox{and}\quad\sum_{k=0}^{\infty}\lambda_{k,l}^2 < \infty,\]
and the errors $\overline{\epsilon}_k, \underline{\epsilon}_k\in\mathbb{R}$ satisfy 
\begin{equation}
\limsup_{k\rightarrow\infty}|\overline{\epsilon}_k|\leq c\quad \mbox{and}\quad \limsup_{k\rightarrow\infty}|\underline{\epsilon}_k|\leq c,
\end{equation}
for some $c\geq 0$, with probability $1$. Then, we have
\begin{equation}
\limsup_{k\rightarrow\infty}\|x_k\|_{\infty} \leq \frac{c}{1-\gamma},
\end{equation}
with probability $1$.
\end{theorem}

\section{Convergence Analysis: Proof of Theorem \ref{thm:main}}\label{sec:analysis}

The proof is built on the following observation: The update of the value function estimate, \eqref{eq:v}, can be written as
\begin{align}\label{eq:finalv}
\hat{v}_{s,k+1}^i = \hat{v}_{s,k}^i + \indicator{s=s_k}\beta_{\#s}\left[T^i(\{\hat{v}^i_{s',k}\}_{s'\in S})[s]-\hat{v}_{s,k}^i + \epsilon_{s,k}^i\right],
\end{align}
where the tracking error $\epsilon_{s,k}^i$ is defined by
\begin{equation}\label{eq:track}
\boxed{\epsilon_{s,k}^i := \inner{\smooth_{s,k}^i}{\hat{q}_{s,k}^i} -
\val^i(\hat{Q}_{s,k}^i)},
\end{equation}
and the operator $T^i:\mathbb{R}^{|S|}\rightarrow \mathbb{R}^{|S|}$ is defined by
\begin{equation}
T^i(\{\hat{v}^i_{s',k}\}_{s'\in S})[s] := \val^i(\hat{Q}^i_{s,k}),
\end{equation}
where $\hat{Q}_{s,k}^i\in[-D,D]^{|A_s^i|\times|A_s^{-i}|}$ corresponding to the global $Q$-function is defined by
\begin{equation}\label{eq:Q}
\hat{Q}_{s,k}^i[a^i,a^{-i}] := r_s^i(a^1,a^2) + \gamma \sum_{s'\in S} p(s'|s,a^1,a^2) \hat{v}_{s',k}^i,\quad \forall i=1,2, 
\end{equation}
and $\val^i:[-D,D]^{|A_s^i|\times|A_s^{-i}|}\rightarrow \mathbb{R}$ is\footnote{Note that $\val^i$ technically also depends on $s$, since $A_s$ depends on $s$. We omit $s$ for notational convenience, and it shall not cause any confusion from the context.} defined by
\begin{equation}\label{eq:val}
\val^i(Q_{s,k}^i) := \max_{\mu^i\in\Delta(A_s^i)}\min_{\mu^{-i}\in\Delta(A_s^{-i})}\left\{(\mu^{i})^TQ_{s,k}^i\mu^{-i}\right\}. 
\end{equation}

The non-expansiveness property of $\val^i(\cdot)$, as shown by \cite{ref:Shapley53}, and the discount factor $\gamma\in (0,1)$ yield that the operator is a contraction, e.g.,
\begin{equation}
\|T^i(v^i)-T^i(\tilde{v}^i)\|_{\infty}\leq \gamma \max_{s\in S}|v^i_s-\tilde{v}^i_s|.
\end{equation}
Denote the unique fixed point of the contraction $T^i$ by $\{v_{s,*}^i\}_{s\in S}$. Then, we have $v_{s,*}^i = T^i(\{v_{s',*}^i\}_{s'\in S})[s]$. Therefore, the update \eqref{eq:finalv} can be written as
\begin{align}\label{eq:finalvdiff}
\hat{v}_{s,k+1}^i&\,-v_{s,*}^i = \hat{v}_{s,k}^i - v_{s,*}^i \nonumber\\
&+ \indicator{s=s_k}\beta_{\#s}\left[T^i(\{\hat{v}^i_{s',k}\}_{s'\in S})[s]-T^i(\{v^i_{s',*}\}_{s'\in S})[s] - (\hat{v}_{s,k}^i - v_{s,*}^i) + \epsilon_{s,k}^i\right],
\end{align}
Based on Proposition \ref{prop:io}, Theorem \ref{thm:bounds} and \eqref{eq:finalv} yield that the asymptotic behavior of the value function estimates can be characterized as follows:
\begin{equation}\label{eq:final}
\limsup_{k\rightarrow\infty}\left|\hat{v}_{s,k}^i - v_{s,*}^i\right|\leq \frac{1}{1-\gamma}\limsup_{k\rightarrow\infty} \epsilon_{s,k}^i
\end{equation}
for all $(i,s)\in\{1,2\}\times S$. The rest of the proof is about characterizing the asymptotic behavior of the tracking error \eqref{eq:track} and showing that 
\begin{equation}\label{eq:trackfinal}
\limsup_{k\rightarrow\infty} \epsilon_{s,k}^i \leq \frac{1}{1-\gamma}\left(\frac{2+\lambda-\lambda\gamma}{1-\lambda\gamma}\right)\max_{s'\in S}\{\log(|A_{s'}^1||A_{s'}^2|)\}\lim_{c\rightarrow\infty}\tau_c,
\end{equation} 
for some $\lambda\in(1,1/\gamma)$.

It is instructive to discuss why the existing results cannot directly address this tracking error's asymptotic behavior. For example, 
there exist several well-established results on convergence properties of learning dynamics in strategic-form games with repeated play for both zero-sum and potential games, e.g., see \cite{ref:Fudenberg09}. The challenge raises since $\hat{Q}_{s,k}^i$ is not time-invariant and it depends on both players' strategies. On the other hand, the existing results to characterize the convergence properties of the classical (single-agent) $Q$-learning is helpful only to obtain \eqref{eq:final} and do not address the tracking error \eqref{eq:track}. Note that if the players are coordinated to play the equilibrium behavior, e.g., as in Shapley's value iteration \citep{ref:Shapley53} or Minimax-Q \citep{littman1994markov}, then the tracking error would be zero by the nature of the updates. However, this would imply that the players are {\em coordinated} to play the equilibrium since
\begin{itemize}
\item Players need to know the zero-sum structure of the game,
\item Players always play the conservative strategy against the worst-case strategy of the opponent and do not attempt to take the best reaction when the opponent is not playing the equilibrium strategy, 
\item Players need to observe the opponent's actions to be able to compute the global $Q$-function associated with the joint actions.  
\end{itemize} 

A two-timescale learning dynamics can address the dependence of the $Q$-function estimate on the strategies and correspondingly address the tracking error. However, there are several challenges especially for radically uncoupled schemes, where players do not observe the opponent's actions:
\begin{itemize}
\item[$i)$] The local $Q$-function estimates for different state and local action pairs can get updated at different frequencies, which poses a challenge for the two-timescale framework to decouple the dynamics at fast and slow timescales. Particularly, the normalization of the step size in the update of the local $Q$-function estimate can ensure that the estimate for each local action gets updated at the same rate in the expectation. However, this is not sufficient since estimates for some local actions can lag behind even the iterates evolving on the slow timescale. 
\item[$ii)$] The $Q$-function estimates may not necessarily sum to zero in general when the players keep track of it independently, i.e., if there is no central coordinator providing it to them. This is important because uncoupled learning dynamics cannot converge to an equilibrium in every class of games, as shown in \cite{hart2003uncoupled}.
\item[$iii)$] The players can keep track of only local $Q$-function since they cannot observe the opponent's action. However, there may not even exist an opponent (mixed) strategy that can lead to the local $Q$-function estimate, i.e., they may not be belief-based, whereas this is not the case if players can observe the opponent's actions to form a belief on the opponent's strategy.
\end{itemize}
In the following, we follow a three-step approach to address these challenges:
\begin{itemize}
\item[$1.$] Decoupling dynamics at the fast timescale by addressing Challenge $i)$.
\item[$2.$] Zooming into the local dynamics (i.e., learning dynamics specific to a single state) at the fast timescale to address Challenges $ii)$ and $iii)$ via a novel Lyapunov function. 
\item[$3.$] Zooming out to the global dynamics (i.e., learning dynamcis across every state) at the slow timescale to characterize the asymptotic behavior of the tracking error \eqref{eq:track}.
\end{itemize}
In the following, we delve into the details of these steps.

\subsection{Decoupling dynamics at the fast timescale} 
Distinct to the radically uncoupled settings, the players can update only the local $Q$-function estimate’s entry specific to the current local action. Although this is an asynchronous update, the normalization makes the evolution of every entry synchronous in the expectation \citep{ref:Leslie05}. To show this, we introduce the stochastic approximation error:
\begin{align}
\omega_{s_k,k}^1[a^1] :=\, &\indicator{a_k^1=a^1}\frac{r_{s}^1(a^{1},a^{2}_k)+\gamma \hat{v}_{s_{k+1},k}^1 - \hat{q}_{s_k,k}^1[a^1]}{\smooth_{k}^1[a^1]} \nonumber\\
&- \E{\indicator{a_k^1=a^1}\frac{r_{s}^1(a^{1},a^{2}_k)+\gamma \hat{v}_{s_{k+1},k}^1 - \hat{q}_{s_k,k}^1[a^1]}{\smooth_{k}^1[a^1]}\,\Big|\,\delta_k},\label{eq:omega}
\end{align}
for all $a^1\in A_s^1$, where $\delta_k:=\{\hat{q}_{s,k}^i,\hat{v}_{s,k}^i\}_{(i,s)\in\{1,2\}\times S}$ includes the iterates at stage $k$, and $\omega_{s_k,k}^2[a^2]$ for $a^2\in A_s^2$ is defined accordingly. The expectation is explicitly given by
\begin{align}
&\E{\indicator{a_k^1=a^1}\frac{r_{s}^1(a^{1},a^{2}_k)+\gamma \hat{v}_{s_{k+1},k}^1 - \hat{q}_{s_k,k}^1[a^1]}{\smooth_{k}^1[a^1]}\,\Big|\,\delta_{k}}\nonumber\\
&\hspace{1.5in}=\smooth_k^1[a^1]\sum_{\tilde{a}^{2}}\smooth_k^{2}[\tilde{a}^{2}]\frac{\hat{Q}_{s_k,k}^1[a^1,\tilde{a}^{2}] - \hat{q}_{s_k,k}^1[a^1]}{\smooth_k^1[a^1]},\label{eq:denom}
\end{align}
where the auxiliary global $Q$-function estimate is as described in \eqref{eq:Q}. Since the denominator disappears in \eqref{eq:denom}, the stochastic approximation error is also given by
\begin{align}
\omega_{s_k,k}^1[a^1] =&\indicator{a_k^1=a^1}\frac{r_{s}^1(a^{1},a^{2}_k)+\gamma \hat{v}_{s_{k+1},k}^1 - \hat{q}_{s_k,k}^1[a^1]}{\smooth_{k}^1[a^1]}\nonumber\\
& - \left(\sum_{\tilde{a}^{2}}\hat{Q}_{s_k,k}^1[a^1,\tilde{a}^{2}] \smooth_{k}^{2}[\tilde{a}^{2}] - \hat{q}_{s_k,k}^1[a^1]\right).
\end{align}
By Proposition \ref{prop:reduce}, we can write \eqref{eq:q} as
\begin{align}\label{eq:sync_q}
\hat{q}_{s_k,k+1}^i = \hat{q}_{s_k,k}^i + \alpha_{\#s_k}\left(\hat{Q}_{s_k,k}^i \smooth_{k}^{-i} - \hat{q}_{s_k,k}^i + \omega_{s_k,k}^i\right),
\end{align}
for $i=1,2$ if $\#s_k\geq C_{s_k}$ and Proposition \ref{prop:io} yields that there exists $\kappa_s\in\mathbb{N}$ such that $\#_{k'}s_k\geq C_{s_k}$ for all $k'\geq \kappa_s$.

Our goal is to characterize the limit set of this discrete-time update for every state. Like \cite{ref:Leslie05}, we can resort to stochastic approximation methods to transform the problem into a tractable continuous-time flow. Distinct to Markov games, we cannot characterize the convergence properties of \eqref{eq:sync_q} for each state separately. By \eqref{eq:Q}, the update \eqref{eq:sync_q} yields that the current state's local Q-function estimate is coupled with any other state's value function estimate. For example, fix an arbitrary state $s$ and take a closer look at how the iterates change in-between two consecutive visits to $s$, denoted by $k$ and $k^{\dagger}$. Since the game does not visit state $s$ until $k^{\dagger}$, we have $\hat{q}_{s,k^{\dagger}}^i = \hat{q}_{s,k+1}^i$ and $\hat{v}_{s,k^{\dagger}}^i = \hat{v}_{s,k+1}^i$ for $i=1,2$, and $\#_{k^{\dagger}}s=\#_ks +1$. In contrast, other states' value function estimates can change depending on the visits to other states at stages within the interval $(k,k^{\dagger})$. Correspondingly, the iterates and the temperature parameter at $k^{\dagger}$ can be written in terms of the iterates at $k$ in the following compact form:
\begin{align}\label{eq:discrete_main}
\begin{bmatrix}\hat{q}_{s,k^{\dagger}}^1\\ \hat{q}_{s,k^{\dagger}}^2\\ \hat{v}_{s,k^{\dagger}}^1\\ \hat{v}_{s,k^{\dagger}}^2 \\ \vdots \\ \hat{v}_{s',k^{\dagger}}^1\\ \hat{v}_{s',k^{\dagger}}^2 \\ \tau_{\#s+1}\end{bmatrix} = \begin{bmatrix}\hat{q}_{s,k}^1\\ \hat{q}_{s,k}^2 \\ \hat{v}_{s,k}^1\\ \hat{v}_{s,k}^2 \\ \vdots \\ \hat{v}_{s',k}^1\\ \hat{v}_{s',k}^2\\\tau_{\#s}\end{bmatrix} + \alpha_{\#s}\left(\begin{bmatrix}\hat{Q}_{s,k}^1 \smoothFunc_{s}^2(\hat{q}_{s,k}^2,\tau_{\#s}) - \hat{q}_{s,k}^1\\ \hat{Q}_{s,k}^2 \smoothFunc_{s}^1(\hat{q}_{s,k}^1,\tau_{\#s}) - \hat{q}_{s,k}^2 \\ 0 \\ 0 \\ \vdots \\ 0 \\ 0 \\ 0 \end{bmatrix} + \begin{bmatrix} \mathbf{0} \\ \mathbf{0} \\ \varepsilon_{s,k^{\dagger}}^1 \\ \varepsilon_{s,k^{\dagger}}^2 \\ \vdots \\ \varepsilon_{s',k^{\dagger}}^1 \\ \varepsilon_{s',k^{\dagger}}^2 \\ \frac{\tau_{\#s+1}-\tau_{\#s}}{\alpha_{\#s}} \end{bmatrix}+\begin{bmatrix}\omega_{s,k}^1\\ \omega_{s,k}^2 \\ 0 \\ 0 \\ \vdots \\ 0 \\ 0 \\ 0\end{bmatrix}\right),
\end{align}
for all $k\geq\kappa_s$, where we define the error terms by
\begin{equation}\label{eq:varepsilon}
\varepsilon_{s',k^{\dagger}}^i := \frac{\hat{v}_{s',k^{\dagger}}^i - \hat{v}_{s',k}^i}{\alpha_{\#s}},\quad\forall (i,s')\in\{1,2\}\times S.
\end{equation} 

Based on Proposition \ref{prop:io}, we can focus on asymptotic convergence properties of \eqref{eq:discrete_main}. Therefore, we are interested in when the convergence properties of \eqref{eq:discrete_main} can be characterized through the following ordinary differential equation (in which the dynamics for $s$ is decoupled from the dynamics for any other state)
\begin{subequations}\label{eq:cont}
\begin{align}
&\frac{dq^1_s(t)}{dt} = \bar{Q}_s^1 \smoothFunc_s^2(q_s^2(t),\bar{\tau}) - q_s^1(t),\\
&\frac{dq^2_s(t)}{dt} = \bar{Q}_s^2 \smoothFunc_s^1(q_s^1(t),\bar{\tau}) - q_s^2(t),
\end{align}
\end{subequations}
for some $\bar{Q}_s^i\in[-D,D]^{|A_s^i|\times|A_s^{-i}|}$ for $i\in\{1,2\}$ and $\bar{\tau}>0$. To this end, we can resort to Theorem \ref{thm:Lyapunov} by showing that the limiting ordinary differential equation of \eqref{eq:cont} is given by
\begin{subequations}\label{eq:cont_big}
\begin{align}
&\frac{dq^1_s(t)}{dt} = Q_s^1(t) \smoothFunc_s^2(q_s^2(t),\tau_s(t)) - q_s^1(t),\label{eq:cont_main1}\\
&\frac{dq^2_s(t)}{dt} = Q_s^2(t) \smoothFunc_s^1(q_s^1(t),\tau_s(t)) - q_s^2(t),\label{eq:cont_main2}\\
&\frac{dv^i_{s'}(t)}{dt} = 0,\quad \forall (i,s')\in \{1,2\}\times S,\label{eq:cont_main3}\\
&\frac{d\tau_s(t)}{dt} = 0,\label{eq:cont_main4}
\end{align}
\end{subequations}
where $Q_s^i[a^i,a^{-i}](t)= r_s^i(a^1,a^2) + \gamma \sum_{s'\in S} p(s'|s,a^1,a^2) v_{s'}^i(t)$ for $i=1,2$. Conditions $i$ (and $ii$) in Theorem \ref{thm:Lyapunov} are satisfied by Assumption \ref{assume:common1} (and Proposition \ref{prop:bound}). Furthermore, the corresponding vector field is Lipschitz continous since it is continously differentiable by \eqref{eq:def_smooth_br_2} and defined over a compact set by Proposition \ref{prop:bound}. The following two lemmas show that the conditions $iv$-$v$ listed in Theorem \ref{thm:Lyapunov} are also satisfied. The proofs of these technical lemmas are provided in Subsection \S\ref{sec:lemmas}.

\begin{lemma}\label{lem:cond3}
Suppose Assumption \ref{assume:common} and either Assumption \ref{assume:first} or \ref{assume:second} hold. Then, the stochastic approximation terms $(\omega_{s,k}^1,\omega_{s,k}^2)$ satisfy \eqref{eq:omegacond} for all $T>0$, $s\in S$, and $i=1,2$.
\end{lemma}

Assumption \ref{assume:first} ensures that the denominator in the update of the local $Q$-function estimate is bounded from below by some non-zero term. On the other hand, Assumption \ref{assume:second} restrains the rate at which the denominator gets close to zero while letting $\lim_{c\rightarrow \infty} \tau_c = 0$.

\begin{lemma}\label{lem:err}
Suppose Assumption \ref{assume:common} and either Assumption \ref{assume:first} or \ref{assume:second} hold. Then, the error terms in \eqref{eq:discrete_main}, $\{\varepsilon^i_{s,k^{\dagger}}\}_{(i,s)\in\{1,2\}\times S}$ and $(\tau_{\#s+1}-\tau_{\#s})/\alpha_{\#s}$, are asymptotically negligible with probability $1$.
\end{lemma}

In the following step, we will zoom into \eqref{eq:cont} and formulate a Lyapunov function to characterize the limit set of not only \eqref{eq:cont} but also the original discrete-time update \eqref{eq:discrete_main}.\footnote{Lyapunov function plays an important role to deduce convergence properties of the discrete-time update via the limiting o.d.e. because the convergence of the limiting o.d.e. does not necessarily imply the convergence of the discrete-time update in general (e.g., see \cite{ref:Benaim99} and \cite{ref:Borkar08}).}  

\subsection{Zooming into local dynamics at the fast timescale}\label{sec:stepII}

In this subsection, we focus only on \eqref{eq:cont} for an arbitrary state $s$, therefore, we drop the subscript $s$ for notational simplicity. For $t\in[0,\infty)$, we focus on the following dynamics
\begin{subequations}\label{eq:cont_q}
\begin{align}
&\frac{dq^1(t)}{dt} = Q^1\smoothFunc^2(q^2(t),\tau) - q^1(t),\\
&\frac{dq^2(t)}{dt} = Q^2\smoothFunc^1(q^1(t),\tau) - q^2(t),
\end{align}
\end{subequations}
with arbitrary initialization of $q^i(0)$ such that $\|q^i(0)\|_{\infty}\leq D$, arbitrary matrices $Q^i$ such that $\|Q^i\|_{\max}\leq D$, and $\tau>0$. The flow \eqref{eq:cont_q} resembles to the local $Q$-functions' evolution in the perturbed best response dynamics:
\begin{subequations}\label{eq:cont_pi}
\begin{align}
&\frac{d\pi^1(t)}{dt} = \smoothFunc^1(Q^1\pi^2(t),\tau) - \pi^1(t),\\
&\frac{d\pi^2(t)}{dt} = \smoothFunc^2(Q^2\pi^1(t),\tau)  - \pi^2(t),
\end{align}
\end{subequations}
where $\pi^i:[0,\infty)\rightarrow \Delta(A^i)$. Indeed, they lead to the same trajectory for $(q^1,q^2)$ if we have $q^1(t) = Q^1\pi^2(t)$ and $q^2(t) = Q^2\pi^1(t)$. However, there may not always exist a strategy, e.g., $\pi^2\in\Delta(A^2)$, such that $q^1(t) = Q^1\pi^2$. If there exists such a strategy, we say $q^1(t)$ is {\em belief-based}, and vice versa. 

We will examine the flow \eqref{eq:cont_q} at a higher-dimensional space to mitigate this issue through 
\begin{subequations}\label{eq:cont_q_higher}
\begin{align}
&\frac{dq^1(t)}{dt} = Q^1\smoothFunc^2(q^2(t),\tau) - q^1(t),\\
&\frac{dq^2(t)}{dt} = Q^2\smoothFunc^1(q^1(t),\tau) - q^2(t),\\
&\frac{d\pi^1(t)}{dt} = \smoothFunc^1(q^1(t),\tau) - \pi^1(t),\\
&\frac{d\pi^2(t)}{dt} = \smoothFunc^2(q^2(t),\tau) - \pi^2(t),
\end{align}
\end{subequations}
where $\pi^i(0)\in\Delta(A^i)$, for $i=1,2$, are initialized arbitrarily. We highlight the differences among \eqref{eq:cont_q}, \eqref{eq:cont_pi}, and \eqref{eq:cont_q_higher}. In \eqref{eq:cont_q_higher}, the dependence between $(q^1,q^2)$ and $(\pi^1,\pi^2)$ is one direction, i.e., the evolution of $(q^1,q^2)$ is as in \eqref{eq:cont_q} and does not depend on $(\pi^1,\pi^2)$. On the contrary, $(\pi^1,\pi^2)$ is not some isolated process as in \eqref{eq:cont_pi}. Its evolution depends on $(q^1,q^2)$ due to $(\smoothFunc^1(q^1(t),\tau),\smoothFunc^2(q^2(t),\tau))$ instead of $(\smoothFunc^1(Q^1\pi^2(t),\tau),\smoothFunc^2(Q^2\pi^1(t),\tau))$. 

We present the following continous and non-negative function as a candidate Lyapunov function for \eqref{eq:cont_q_higher}:
\begin{align}\label{eq:Lyapunov}
V(q^1,q^2,\pi^1,\pi^2) := &\left[\sum_{i=1,2}\max_{\mu\in\Delta(A^i)}\{\inner{\mu}{q^i} + \tau \nu^i(\mu)\} - \lambda \zeta\right]_++\sum_{i=1,2}\|q^i- Q^i\pi^{-i}\|^2,
\end{align}
where we define $[g(t)]_+ := \max\{g(t),0\}$ for a given function $g(\cdot)$, the auxiliary parameter $\lambda \in (1,1/\gamma)$ is arbitrary, and we define
\begin{equation}\label{eq:zeta}
\zeta:= \|Q^1+(Q^2)^T\|_{\max} + \tau\log(|A^1||A^2|). 
\end{equation}
Note that $\zeta\geq 0$ depends on $Q^1,Q^2$, and $\tau$ implicitly, and it is small when the auxiliary game is close to zero-sum and the temperature parameter is close to zero. The arbitrary parameter $\lambda>1$ plays an important role in ensuring that the set $\{(q^1,q^2,\pi^1,\pi^2): V(q^1,q^2,\pi^1,\pi^2)=0\}$ is a global attractor for the flow \eqref{eq:cont_q_higher}. Furthermore, the condition $\lambda\gamma\in(0,1)$ will play an important role when we zoom out to the global dynamics in Subsection \S\ref{sec:step3}.

Before validating $V(\cdot)$ as a Lyapunov function, let us highlight its differences from other Lyapunov functions used for the best response dynamics with or without perturbation. For example, \cite{ref:Hofbauer05} provided a Lyapunov function for the perturbed best response dynamics in zero-sum games and showed that such dynamics converge to a \emph{Nash distribution}  for any smooth function and any positive temperature parameter. However, we must consider arbitrary $Q^1$ and $Q^2$, which implies that $Q^1+(Q^2)^T$ may not be a \emph{zero matrix} in general. In other words, the underlying game is not necessarily zero-sum. Therefore, we need to consider this deviation in our candidate function.

On the other hand, \cite{ref:Sayin20} provided a Lyapunov function for the best response dynamics in games beyond zero-sum and showed that such dynamics converge to a bounded set with diameter depending on its deviation from a zero-sum game. Therefore, our candidate \eqref{eq:Lyapunov} has a similar flavor with the one in \cite{ref:Sayin20} while addressing also the perturbation and the issue induced by not being belief-based. For example, $V(q_*^1,q_*^2,\pi_*^1,\pi_*^2) = 0$ implies that
\begin{align}
\sum_{i=1,2}\max_{\mu\in\Delta(A^i)}\{\inner{\mu}{q_*^i} + \tau \nu^i(\mu)\} \leq \lambda \zeta,
\end{align}
which yields that
\begin{align}
\sum_{i=1,2}\inner{q_*^i}{\smoothFunc^i(q_*^i,\tau)} \leq \lambda \zeta, 
\end{align}
since the smooth functions $\nu^i(\cdot)$ are non-negative. Recall that the players update their value function estimates, e.g., $\hat{v}_{s,k}^i$, towards $\inner{\hat{q}_{s,k}^i}{\smoothFunc^i_s(\hat{q}_{s,k}^i,\tau_{\#s})}$. Therefore, such an upper bound plays an important role in chacterizing the convergence properties of the sum $\hat{v}_{s,k}^1+\hat{v}_{s,k}^2$ and addressing the deviation from zero-sum settings. Furthermore, $V(q_*^1,q_*^2,\pi_*^1,\pi_*^2) = 0$ also yields that  $q^i_* = Q^i\pi_*^{-i}$, i.e., $(q_*^1,q_*^2)$ are belief-based. It is also instructive to note that $(q_*^1,q_*^2,\pi_*^1,\pi_*^2)$ is not necessarily an equilibrium point of the flow \eqref{eq:cont_q_higher}. Indeed, such a Lyapunov function does not exist because the flow \eqref{eq:cont_q_higher} (and best response dynamics) is not globally asymptotically stable for arbitrary matrices $(Q^1,Q^2)$. 

The following lemma shows that the non-negative $V(\cdot)$ is a Lyapunov function for the flow \eqref{eq:cont_q_higher} and its proof is provided in Subsection \S\ref{sec:lemmas}.

\begin{lemma}\label{lem:Lyapunov}
Consider any trajecttory of \eqref{eq:cont_q_higher} and let $x(t):=(q^1(t),q^2(t),\pi^1(t),\pi^2(t))$. Then the candidate function $V(\cdot)$, as described in \eqref{eq:Lyapunov}, satisfies 
\begin{itemize}
\item $V(x(t')) < V(x(t))$ for all $t'>t$ if $V(x(t))>0$,
\item $V(x(t')) = 0$ for all $t'>t$ if $V(x(t))=0$.
\end{itemize} 
\end{lemma}

Based on Lemma \ref{lem:Lyapunov}, we can characterize the convergence properties of the discrete-time update \eqref{eq:discrete_main}. There is a sequence of beliefs $\{\hat{\pi}_{s,k}^j\in\Delta(A_s^j)\}_{k\geq 0}$ for the sequence $\{\hat{q}_{s,k}^i\}_{k\geq 0}$ and it evolves according to 
\begin{align}\label{eq:belief}
\hat{\pi}_{s,k+1}^j = \left\{\begin{array}{ll}
\hat{\pi}_{s,k}^j + \alpha_{\#s}\left(\smooth_k^j - \hat{\pi}_{s,k}^j\right)&\mbox{if } s=s_k\\
\hat{\pi}_{s,k}^j & \mbox{o.w.}
\end{array}\right.
\end{align} 
with some arbitrary initialization, and satisfies
\begin{equation}\label{eq:beliefbased}
\lim_{k\rightarrow \infty} \|\hat{q}_{s,k}^i - \hat{Q}_{s,k}^i \hat{\pi}_{s,k}^j\|^2 = 0,
\end{equation}
where $j\neq i$. Denote $\bar{Q}_{s,k}:= \bar{Q}_{s,k}^1+(\bar{Q}_{s,k}^2)^T$. Then, Lemma \ref{lem:Lyapunov} yields that
\begin{align*}
\lim_{k\rightarrow\infty} \left[\sum_{i=1,2} \left(\inner{\hat{q}_{s,k}^i}{\smooth_{k}^i} + \tau_{\#s} \nu_s^i(\smooth_{k}^i)\right) - \lambda \left(\|\bar{Q}_{s,k}\|_{\max} + \tau_{\#s}\log(|A_s^1||A_s^2|)\right)\right]_+ = 0,
\end{align*}
which implies that there exists $\{\overline{e}_{s,k}\geq 0\}_{k\geq 0}$ and $\lim_{k\rightarrow \infty} \overline{e}_{s,k} = 0$ such that
\begin{equation}\label{eq:upper}
\sum_{i=1,2} \left(\inner{\hat{q}_{s,k}^i}{\smooth_{k}^i} + \tau_{\#s} \nu_s^i(\smooth_{k}^i)\right) \leq \lambda \left(\|\bar{Q}_{s,k}\|_{\max} + \tau_{\#s}\log(|A_s^1||A_s^2|)\right) + \overline{e}_{s,k}, \quad \forall k\geq 0.
\end{equation}

In the following, we characterize the convergence properties of the value function estimates based on \eqref{eq:beliefbased} and \eqref{eq:upper}, respectively, showing that the local $Q$-function estimates are asymptotically belief-based and characterizing an upper bound on the sum of the (perturbed) values.

\subsection{Zooming out to global dynamics at the slow timescale}\label{sec:step3}

Next, we focus on the evolution of the value function estimates. To this end, we first consider how the sum of the players' value function estimates specific to state $s$ (denoted by $\bar{v}_{s,k}:=\hat{v}_{s,k}^1+\hat{v}_{s,k}^2$) evolves:
\begin{equation}\label{eq:sum}
\bar{v}_{s,k+1} = \left\{\begin{array}{ll}
\bar{v}_{s,k} + \beta_{\#s}\left[\sum_{i=1,2}\inner{\smooth_{k}^i}{\hat{q}_{s,k}^i} - \bar{v}_{s,k}\right] & \mbox{if } s=s_k\\
\bar{v}_{s,k} & \mbox{o.w.}
\end{array}\right.
\end{equation}
We can view \eqref{eq:sum} as the sum $\bar{v}_{s,k}$ moving toward (or tracking) the target $\sum_{i=1,2}\inner{\smooth_{k}^i}{\hat{q}_{s,k}^i}$. The target is bounded from above by
\begin{equation}\label{eq:upper2}
\sum_{i=1,2}\inner{\smooth_{k}^i}{\hat{q}_{s,k}^i} \leq \lambda \|\bar{Q}_{s,k}\|_{\max} + \lambda\tau_{\#s}\log(|A_s^1||A_s^2|) + \overline{e}_{s,k},\quad\forall k\geq 0,
\end{equation}
by \eqref{eq:upper}. We can also bound the target from below by using the smooth best response definition and \eqref{eq:beliefbased} as follows:
\begin{align}\label{eq:lower}
\sum_{i=1,2}\left(\inner{\smooth_{k}^i}{\hat{q}_{s,k}^i} + \tau_{\#s}\nu_s^i(\smooth_{s,k}^i)\right) \geq (\hat{\pi}_{s,k}^1)^T\bar{Q}_{s,k}\hat{\pi}_{s,k}^2 + \tau_{\#s}\sum_{i=1,2}\nu_s^i(\hat{\pi}_{s,k}^i) + \underline{e}_{s,k},\quad \forall k\geq 0,
\end{align}
where the error term is given by
\begin{equation}
\underline{e}_{s,k} := \sum_{i=1,2}(\hat{\pi}_{s,k}^i)^T(\hat{q}_{s,k}^i-\hat{Q}_{s,k}^i\hat{\pi}_{s,k}^{-i})
\end{equation}
and it is asymptotically negligible by \eqref{eq:beliefbased}. Since $\hat{\pi}_{s,k}^i\in\Delta(A_s^i)$, we obtain
\begin{align}
\sum_{i=1,2}\inner{\smooth_{k}^i}{\hat{q}_{s,k}^i} &\geq -\|\bar{Q}_{s,k}\|_{\max} +\tau_{\#s}\sum_{i=1,2}(\nu_s^i(\hat{\pi}_{s,k}^i)-\nu_s^i(\smooth_k^i)) + \underline{e}_{s,k}\\
&\geq -\lambda\|\bar{Q}_{s,k}\|_{\max} - \lambda\tau_{\#s}\log(|A_s^1||A_s^2|) + \underline{e}_{s,k},\label{eq:lower2}
\end{align}
where the last inequality follows since $\lambda>1$ and $\nu_s^i:\Delta(A_s^i)\rightarrow [0,\log(|A_s^i|)]$. 

Based on the fact that $r_s^1(a^1,a^2)+r_s^2(a^1,a^2)=0$ for all $(a^1,a^2)$, we can formulate a bound on $\|\bar{Q}_{s,k}\|_{\max}$ from above in terms of $\{\bar{v}_{s',k}\}_{s'\in S}$ as follows:
\begin{align}
\|\bar{Q}_{s,k}\|_{\max} & = \max_{(a^1,a^2)} \left|r_s^1(a^1,a^2)+r_s^2(a^1,a^2)+\gamma \sum_{s'\in S} p(s'|s,a^1,a^2) \bar{v}_{s',k}\right|\nonumber\\
&\leq \gamma \max_{s'\in S} |\bar{v}_{s',k}|.\label{eq:upper3}
\end{align} 
Combining \eqref{eq:upper2}, \eqref{eq:lower2}, and \eqref{eq:upper3}, we obtain
\begin{align}\label{eq:lowupp}
-\lambda\gamma \max_{s'\in S} |\bar{v}_{s',k}| - \lambda\tau_{\#s}\log(|A_s^1||A_s^2|) + &\,\underline{e}_{s,k} \leq \sum_{i=1,2}\inner{\smooth_{k}^i}{\hat{q}_{s,k}^i}  \nonumber\\
&\leq \lambda\gamma \max_{s'\in S} |\bar{v}_{s',k}| + \lambda\tau_{\#s}\log(|A_s^1||A_s^2|)+\overline{e}_{s,k},
\end{align}
for all $s\in S$ and $k\geq 0$, with some asymptotically negligible error terms $\underline{e}_{s,k}$ and $\overline{e}_{s,k}$. The condition $\lambda\gamma\in(0,1)$ yields that the target in \eqref{eq:sum} shrinks in absolute value as $k\rightarrow \infty$. Based on \eqref{eq:sum} and Theorem \ref{thm:bounds}, we obtain
\begin{align}\label{eq:sumlim}
\limsup_{k\rightarrow\infty} \max_{s\in S}|\bar{v}_{s,k}| \leq \frac{1}{1-\lambda\gamma}\lambda\xi\lim_{c\rightarrow\infty}\tau_c,
\end{align}
where $\xi := \max_{s\in S}\{\log(|A_s^1||A_s^2|)\}$ since $\#s\rightarrow\infty$ as $k\rightarrow\infty$ with probability $1$ by Proposition \ref{prop:io}. Therefore, the auxiliary games get close to (or become) zero-sum asymptotically like the two-timescale fictitious play in \cite{ref:Sayin20} but with a radically uncoupled scheme. 

The next and last step is about characterizing the asymptotic behavior of the tracking error \eqref{eq:track}:
\begin{align}
&\limsup_{k\rightarrow\infty}\left|\inner{\smooth_{s,k}^1}{\hat{q}_{s,k}^1} -
\val^1(\hat{Q}_{s,k}^1)\right| \nn \\
&\hspace{0.5in}\stackrel{(a)}{\leq}\limsup_{k\rightarrow\infty}\left|\max_{\mu^1}\left\{(\mu^1)^T\hat{Q}_{s,k}^1\hat{\pi}_{s,k}^2\right\} -
\val^1(\hat{Q}_{s,k}^1)\right|\nonumber \nn \\
&\hspace{0.7in}+ \limsup_{k\rightarrow\infty} \left|\max_{\mu^1}\left\{(\mu^1)^T\hat{Q}_{s,k}^1\hat{\pi}_{s,k}^2\right\}-\inner{\smooth_{s,k}^1}{\hat{q}_{s,k}^1}\right| \nn\\
&\hspace{0.5in}\stackrel{(b)}{\leq}\limsup_{k\rightarrow\infty}\left|\max_{\mu^1}\left\{(\mu^1)^T\hat{Q}_{s,k}^1\hat{\pi}_{s,k}^2\right\} -
\val^1(\hat{Q}_{s,k}^1)\right| + \xi\lim_{c\rightarrow\infty}\tau_c\nn\\
&\hspace{0.5in}\stackrel{(c)}{\leq} \limsup_{k\rightarrow\infty}\left|\max_{\mu^1}\left\{(\mu^1)^T\hat{Q}_{s,k}^1\hat{\pi}_{s,k}^2\right\} -
\min_{\mu^2}\left\{(\hat{\pi}_{s,k}^1)^T\hat{Q}_{s,k}^1\mu^2\right\}\right|+ \xi\lim_{c\rightarrow\infty}\tau_c\nn\\
&\hspace{0.5in}\stackrel{(d)}{\leq}\limsup_{k\rightarrow\infty}\left|\max_{\mu^2}\left\{(\mu^2)^T\hat{Q}_{s,k}^2\hat{\pi}_{s,k}^1\right\} +
\min_{\mu^2}\left\{(\mu^2)^T(\hat{Q}_{s,k}^1)^T\hat{\pi}_{s,k}^1\right\}\right|\nonumber \nn\\
&\hspace{0.7in}+ \limsup_{k\rightarrow\infty}\left|\sum_{i=1,2} \max_{\mu^i}\{\inner{\mu^i}{\hat{Q}_{s,k}^i\hat{\pi}_{s,k}^{-i}}\}\right|+ \xi\lim_{c\rightarrow\infty}\tau_c\nn\\
&\hspace{0.5in}\stackrel{(e)}{\leq} \limsup_{k\rightarrow\infty}\|\bar{Q}_{s,k}\|_{\max}+ \limsup_{k\rightarrow\infty}\left|\sum_{i=1,2} \max_{\mu^i}\{\inner{\mu^i}{\hat{Q}_{s,k}^i\hat{\pi}_{s,k}^{-i}}\}\right|+ \xi\lim_{c\rightarrow\infty}\tau_c\nn\\
&\hspace{0.5in}\stackrel{(f)}{\leq} \gamma \limsup_{k\rightarrow\infty}\max_{s'\in S}|\bar{v}_{s',k}|+ \limsup_{k\rightarrow\infty}\left|\sum_{i=1,2} \max_{\mu^i}\{\inner{\mu^i}{\hat{Q}_{s,k}^i\hat{\pi}_{s,k}^{-i}}\}\right|+ \xi\lim_{c\rightarrow\infty}\tau_c\nn\\
&\hspace{0.5in}\stackrel{(g)}{\leq} \gamma \limsup_{k\rightarrow\infty}\max_{s'\in S}|\bar{v}_{s',k}|+ \limsup_{k\rightarrow\infty}\left|\sum_{i=1,2} \max_{\mu^i}\{\inner{\mu^i}{\hat{q}_{s,k}^i}\}\right|+ \xi\lim_{c\rightarrow\infty}\tau_c\nn\\
&\hspace{0.5in}\stackrel{(h)}{\leq} \gamma \limsup_{k\rightarrow\infty}\max_{s'\in S}|\bar{v}_{s',k}|+\limsup_{k\rightarrow\infty}\left|\sum_{i=1,2}\inner{\smooth_k^i}{\hat{q}_{s,k}^i}\right| +  2\xi\lim_{c\rightarrow\infty}\tau_c\nn\\
&\hspace{0.5in}\stackrel{(i)}{\leq} (\lambda\gamma + \gamma)\limsup_{k\rightarrow\infty} \max_{s\in S}|\bar{v}_{s,k}| + (\lambda+2)\xi\lim_{c\rightarrow\infty}\tau_c\nn\\
&\hspace{0.5in}\stackrel{(j)}{\leq} \left(\frac{\lambda(\lambda\gamma + \gamma)}{1-\lambda\gamma} + (\lambda+2)\right)\xi \lim_{c\rightarrow\infty}\tau_c\nn\\
&\hspace{0.5in}=\left(\frac{2+\lambda-\lambda\gamma}{1-\lambda\gamma}\right)\xi\lim_{c\rightarrow\infty}\tau_c.\label{eq:trackfinal}
\end{align}
Particularly, $(a)$ follows from triangle inequality; $(b)$ follows since 
\begin{align}\label{eq:bestsmoothbound1}
\inner{\smooth_k^i}{\hat{q}_{s,k}^i} + \tau_{\#s} \log(|A_s^i|) \geq \max_{\mu^i\in\Delta(A_s^i)}\{\inner{\mu^i}{\hat{q}_{s,k}^i}\} \geq \inner{\smooth_k^i}{\hat{q}_{s,k}^i},
\end{align} 
by definition of best response and smooth best response and since $\hat{q}_{s,k}^i$ is asymptotically belief-based and $\max\{\cdot\}$ is a continuous operator; $(c)$ follows from the fact that
\begin{equation}
\max_{\mu^1}\left\{(\mu^1)^T\hat{Q}_{s,k}^1\hat{\pi}_{s,k}^2\right\} \geq \val^1(\hat{Q}_{s,k}^1) \geq \min_{\mu^2}\left\{(\hat{\pi}^1_{s,k})^T\hat{Q}_{s,k}^1\mu^2\right\};
\end{equation}
$(d)$ follows from the triangle inequality; $(e)$ follows since we have 
\begin{align}
&\left|\max_{\mu^2}\left\{(\mu^2)^T\hat{Q}_{s,k}^2\hat{\pi}_{s,k}^1\right\} +
\min_{\mu^2}\left\{(\mu^2)^T(\hat{Q}_{s,k}^1)^T\hat{\pi}_{s,k}^1\right\}\right| \nonumber\\
&\hspace{1in}= \left|\max_{\mu^2}\left\{(\mu^2)^T\hat{Q}_{s,k}^2\hat{\pi}_{s,k}^1\right\} - 
\max_{\mu^2}\left\{(\mu^2)^T(-\hat{Q}_{s,k}^1)^T\hat{\pi}_{s,k}^1\right\}\right|\label{eq:difff}\\
&\hspace{1in}\leq \|\bar{Q}_{s,k}\|_{\max},
\end{align}
where \eqref{eq:difff} corresponds to the difference between the maximum values player $2$ would get in the scenarios in which it has the payoff matrices $\hat{Q}_{s,k}^2$ versus $(-\hat{Q}_{s,k}^1)$ in an auxiliary strategic-form game, given that the opponent's play is fixed, and this difference is bounded from above by $\|\hat{Q}_{s,k}^2-(-\hat{Q}_{s,k}^1)\|_{\max}=\|\bar{Q}_{s,k}\|_{\max}$; $(f)$ follows from \eqref{eq:upper3}; $(g)$ follows since $\hat{q}_{s,k}^i$ is asymptotically belief-based; $(h)$ follows from \eqref{eq:bestsmoothbound1}; $(i)$ follows since \eqref{eq:lowupp} yields that
\begin{equation}
\limsup_{k\rightarrow\infty}\left|\sum_{i=1,2}\inner{\smooth_{k}^i}{\hat{q}_{s,k}^i}\right| \leq \lambda\gamma \limsup_{k\rightarrow\infty} \max_{s\in S}|\bar{v}_{s,k}| + \lambda\xi\lim_{c\rightarrow\infty}\tau_c;
\end{equation}
and finally $(j)$ follows from \eqref{eq:sumlim}.  This completes the first part of the result on the asymptotic behavior of $\{\hat{v}_{s,k}^i\}_{k\geq 0}$. 

On the other hand, the weighted time-average of smoothed best responses corresponds to $\hat{\pi}_{s,k}^i$, as described iteratively in \eqref{eq:belief}. Note that $v_{\pi_*}^i(s) = \val^i(Q_{\pi_*}^i(s,\cdot))$ by its definition. Therefore, for any $s$, we define $v_{*,\pi^{-i}}^{-i}(s):=\min\limits_{\pi^{i}}v_{\pi^i,\pi^{-i}}^{-i}(s)=-\max\limits_{\pi^{i}}v_{\pi^i,\pi^{-i}}^{i}(s)=:-v_{*,\pi^{-i}}^{i}(s)$. Note that the convention is that for agent $i$ and her value function $v^i_{\pi^i,\pi^{-i}}$, we use $v^i_{*,\pi^{-i}}$ to denote the $\max$ over her own strategy $\pi^i$, and $v^i_{\pi^i,*}$ to denote the $\min$ over her opponent's strategy $\pi^{-i}$. In other words, agent $i$ always maximizes her value, while her opponent always minimizes it. Also note that for fixed strategy of one player, the problem is a Markov decision process, which always admits some maximizing/minimizing strategy for all $s$, i.e., these best-response values are well-defined. Finally, for these value functions, one can define the corresponding $Q$-functions  satisfying the following Bellman equations: 
\begin{align*}
Q_{*,\pi^{-i}}^i(s,a^1,a^2) &=\, r_{s}^i(a^1,a^2)+ \gamma \sum_{s'\in S} v_{*,\pi^{-i}}^i(s')p(s'|s,a^1,a^2), \quad \forall (s,a^1,a^2)\\
v_{*,\pi^{-i}}^i(s) &=\max_{\mu}~\mathbb{E}_{a^{i}\sim \mu,~a^{-i}\sim \pi^{-i}_s}\Big[Q_{*,\pi^{-i}}^i(s,a^1,a^2)\Big], \quad \forall~s,
\end{align*}
and 
\begin{align*}
Q_{\pi_*}^i(s,a^1,a^2) &=\, r_{s}^i(a^1,a^2)+ \gamma \sum_{s'\in S} v_{\pi_*}^i(s')p(s'|s,a^1,a^2), \quad \forall (s,a^1,a^2)\\
v_{\pi_*}^i(s) &=\max_{\mu}\min_{\nu}~\mathbb{E}_{a^{i}\sim \mu,~a^{-i}\sim \nu}\Big[Q_{\pi_*}^i(s,a^1,a^2)\Big], \quad \forall~s. 
\end{align*}
Other quantities can be defined similarly.

Therefore, we have 
\begin{align} 
0&\leq v_{\pi_*}^{-i}(s)-v_{*,\hat{\pi}_{k}^{-i}}^{-i}(s)=-\max_{\mu^i}\,(\mu^i)^TQ_{\pi_*}^i(s,\cdot)\pi_{s,*}^{-i}+\max_{\mu^i}\,(\mu^i)^TQ_{*,\hat{\pi}_{k}^{-i}}^i(s,\cdot)\hat{\pi}_{s,k}^{-i}\label{equ:approx_NE_1}\\
&\leq \Big|\max_{\mu^i}\,(\mu^i)^TQ_{*,\hat{\pi}_{k}^{-i}}^i(s,\cdot)\hat{\pi}_{s,k}^{-i}-\max_{\mu^i}\,(\mu^i)^TQ_{\pi_*}^i(s,\cdot)\hat{\pi}_{s,k}^{-i}\Big|\notag\\
&\quad\quad+\Big|\max_{\mu^i}\,(\mu^i)^TQ_{\pi_*}^i(s,\cdot)\hat{\pi}_{s,k}^{-i}-\max_{\mu^i}\,(\mu^i)^TQ_{\pi_*}^i(s,\cdot)\pi_{s,*}^{-i}\Big|\label{equ:approx_NE_2}\\
&\leq \gamma\cdot\max_{s\in S}\big|v_{*,\hat{\pi}_{k}^{-i}}^i(s)-v_{\pi_*}^i(s)\big|+\Big|\val^i(\hat{Q}_{s,k}^i)-\max_{\mu^i}\,(\mu^i)^TQ_{\pi_*}^i(s,\cdot)\pi_{s,*}^{-i}\Big|\notag\\
&\quad\quad +\Big|\max_{\mu^i}\,(\mu^i)^TQ_{\pi_*}^i(s,\cdot)\hat{\pi}_{s,k}^{-i}-\val^i(\hat{Q}_{s,k}^i)\Big|\label{equ:approx_NE_3}\\
&\leq \gamma\cdot\max_{s\in S}\big|v_{\pi_*}^{-i}(s)-v_{*,\hat{\pi}_{k}^{-i}}^{-i}(s)\big|+\Big|\val^i(\hat{Q}_{s,k}^i)-\max_{\mu^i}\,(\mu^i)^TQ_{\pi_*}^i(s,\cdot)\pi_{s,*}^{-i}\Big|\notag\\
&\quad\quad +\Big|\max_{\mu^i}\,(\mu^i)^TQ_{\pi_*}^i(s,\cdot)\hat{\pi}_{s,k}^{-i}-\max_{\mu^i}\,(\mu^i)^T\hat{Q}_{s,k}^i(s,\cdot)\hat{\pi}_{s,k}^{-i}\Big|\notag\\
&\quad\quad+\Big|\max_{\mu^i}\,(\mu^i)^T\hat{Q}_{s,k}^i(s,\cdot)\hat{\pi}_{s,k}^{-i}-\val^i(\hat{Q}_{s,k}^i)\Big|,\label{equ:approx_NE_4}
\end{align}
where \eqref{equ:approx_NE_1} is due to one-step Bellman equation and the zero-sum structure of the underlying game, \eqref{equ:approx_NE_2} follows by inserting $\max_{\mu^i}\,(\mu^i)^TQ_{\pi_*}^i(s,\cdot)\hat{\pi}_{s,k}^{-i}$, \eqref{equ:approx_NE_3} follows by the fact that 
\begin{align*}
&\Big|\max_{\mu^i}\,(\mu^i)^TQ_{*,\hat{\pi}_{k}^{-i}}^i(s,\cdot)\hat{\pi}_{s,k}^{-i}-\max_{\mu^i}\,(\mu^i)^TQ_{\pi_*}^i(s,\cdot)\hat{\pi}_{s,k}^{-i}\Big|\\
&\quad\leq \max_{\mu^i}~\Big|(\mu^i)^TQ_{*,\hat{\pi}_{k}^{-i}}^i(s,\cdot)\hat{\pi}_{s,k}^{-i}-(\mu^i)^TQ_{\pi_*}^i(s,\cdot)\hat{\pi}_{s,k}^{-i}\Big|\leq \gamma \cdot\max_{s\in S}\big|v_{*,\hat{\pi}_{k}^{-i}}^i(s)-v_{\pi_*}^i(s)\big|
\end{align*}
and by inserting $\val^i(\hat{Q}_{s,k}^i)$, and finally, \eqref{equ:approx_NE_4} follows by inserting $\max_{\mu^i}\,(\mu^i)^T\hat{Q}_{s,k}^i(s,\cdot)\hat{\pi}_{s,k}^{-i}$, and by the zero-sum property of the values.  The last three terms in \eqref{equ:approx_NE_4} can be further bounded as follows: 1) by definitions, we have 
\begin{align}
&\Big|\val^i(\hat{Q}_{s,k}^i)-\max_{\mu^i}\,(\mu^i)^TQ_{\pi_*}^i(s,\cdot)\pi_{s,*}^{-i}\Big|
\leq \big\|\hat{Q}_{s,k}^i-Q_{\pi_*}^i(s,\cdot)\big\|_{\max}\notag\\
&\quad\leq \gamma \max_{s'\in S} |v_{\pi_*}^i(s') - \hat{v}_{s',k}^i|; \label{equ:approx_NE_5}\\
&\Big|\max_{\mu^i}\,(\mu^i)^TQ_{\pi_*}^i(s,\cdot)\hat{\pi}_{s,k}^{-i}-\max_{\mu^i}\,(\mu^i)^T\hat{Q}_{s,k}^i(s,\cdot)\hat{\pi}_{s,k}^{-i}\Big|\notag\\
&\quad\leq \big\|\hat{Q}_{s,k}^i-Q_{\pi_*}^i(s,\cdot)\big\|_{\max}\leq \gamma \max_{s'\in S} |v_{\pi_*}^i(s') - \hat{v}_{s',k}^i|; \label{equ:approx_NE_6}
\end{align}
2) by $(c)$-$(i)$ in \eqref{eq:trackfinal}, we have 
\begin{align}
\limsup_{k\rightarrow\infty} \left|\max_{\mu^i} \,(\mu^i)^T\hat{Q}_{s,k}^i\hat{\pi}_{s,k}^{-i} - \val^i(\hat{Q}_{s,k}^i)\right| \leq \left(\frac{1+\lambda}{1-\lambda \gamma}\right)\xi\lim_{c\rightarrow\infty} \tau_c.\label{equ:approx_NE_7}
\end{align}
Hence, combining \eqref{equ:approx_NE_4}-\eqref{equ:approx_NE_7}, we have  
\begin{align}\label{equ:approx_NE_8}
&\limsup_{k\rightarrow\infty}\max_{s\in S}\big|v_{\pi_*}^{-i}(s)-v_{*,\hat{\pi}_{k}^{-i}}^{-i}(s)\big|\leq \frac{1}{1-\gamma}\cdot \Bigg[2\gamma \limsup_{k\rightarrow\infty}\max_{s'\in S} |v_{\pi_*}^i(s') - \hat{v}_{s',k}^i|+\left(\frac{1+\lambda}{1-\lambda \gamma}\right)\xi\lim_{c\rightarrow\infty} \tau_c\Bigg]\notag\\
& \leq \frac{1}{1-\gamma}\cdot \Bigg[2\gamma \xi g(\gamma)\lim_{c\rightarrow\infty} \tau_c+\left(\frac{1+\lambda}{1-\lambda \gamma}\right)\xi\lim_{c\rightarrow\infty} \tau_c\Bigg].
\end{align}
Finally, notice the fact that 
\begin{align}\label{equ:approx_NE_9} 
&0\leq v_{*,\hat{\pi}_{k}^{-i}}^{i}(s)-v_{\hat{\pi}_{k}^{i},\hat{\pi}_{k}^{-i}}^{i}(s)\leq  v_{*,\hat{\pi}_{k}^{-i}}^{i}(s)-v_{\hat{\pi}_{k}^{i},*}^{i}(s)=-v_{*,\hat{\pi}_{k}^{-i}}^{-i}(s)+v_{\hat{\pi}_{k}^{i},*}^{-i}(s)
\\
&\quad =-v_{\hat{\pi}_{k}^{i},*}^{i}(s)+v_{\pi_*}^{i}(s)+v_{\pi_*}^{-i}(s)-v_{*,\hat{\pi}_{k}^{-i}}^{-i}(s)\notag\\
&\quad\leq \max_{s\in S}\big|v_{\pi_*}^{i}(s)-v_{\hat{\pi}_{k}^{i},*}^{i}(s)\big|+\max_{s\in S}\big|v_{\pi_*}^{-i}(s)-v_{*,\hat{\pi}_{k}^{-i}}^{-i}(s)\big|_{\max},\notag
\end{align}
where the last inequality follows from \eqref{equ:approx_NE_8} and its counterpart by switching the role of $-i$ and $i$ there in. 
Combined with the definition of $\epsilon$-Nash equilibrium with \eqref{equ:approx_NE_8}-\eqref{equ:approx_NE_9}, we complete the proof. 

\subsection{Proofs of Technical Lemmas \ref{lem:cond3}-\ref{lem:Lyapunov}}\label{sec:lemmas}

\begin{lemma*}{Lemma \ref{lem:cond3}}
Suppose Assumption \ref{assume:common} and either Assumption \ref{assume:first} or \ref{assume:second} hold. Then, the stochastic approximation terms $(\omega_{s,k}^1,\omega_{s,k}^2)$ satisfy \eqref{eq:omegacond} for all $T>0$, $s\in S$, and $i=1,2$.
\end{lemma*}

\begin{myproof}
By \eqref{eq:omega}, the stochastic approximation term $\omega_{s,k}^1[a^1]$ (and $\omega_{s,k}^2$) can be written as
\begin{equation}
\omega_{s,k}^1[a^1] = \frac{\tilde{\omega}_{s,k}^1[a^1]}{\smooth_{k}^1[a^1]},\quad\forall a^1\in A_s^1,
\end{equation}
where we define 
\begin{align}
\tilde{\omega}_{s,k}^1[a^1] :=&\, \indicator{a_k^1=a^1}(r_{s}^1(a^{1},a^{2}_k) + \gamma \hat{v}_{s_{k+1},k}^1 - \hat{q}_{s,k}^1[a^1]) \nonumber\\
&- \E{\indicator{a_k^1=a^1}(r_{s}^1(a^{1},a^{2}_k) + \gamma \hat{v}_{s_{k+1},k}^1 - \hat{q}_{s,k}^1[a^1])\,\Big|\,\delta_k}\\
=&\, \indicator{a_k^1=a^1}(r_{s}^1(a^{1},a^{2}_k) + \gamma \hat{v}_{s_{k+1},k}^1)\nonumber \\
& - \E{\indicator{a_k^1=a^1}(r_{s}^1(a^{1},a^{2}_k) + \gamma \hat{v}_{s_{k+1},k}^1)\,\Big|\,\delta_k},
\end{align}
which is a square-integrable Martingale difference sequence since the iterates remain bounded by Proposition \ref{prop:bound}. 
Then, the proof follows from the proof of \cite[Proposition 4.2]{ref:Benaim99} by substituting the step size $\alpha_{\#s}$ with $\frac{\alpha_{\#s}}{\smooth_{k}^i[a^i]}$ and showing that
\begin{equation}\label{eq:limsum}
\sum_{k\geq 0:s=s_k} \left(\frac{\alpha_{\#s}}{\smooth_{k}^i[a^i]}\right)^2 < \infty,\quad\forall a^i\in A_s^i\mbox{ and }s\in S.
\end{equation} 

If Assumption \ref{assume:first} holds, then the analytical form of $\smooth_k^i[a^i]$, as described in \eqref{eq:def_smooth_br_2}, yields that $\smooth_k^i[a^i]\geq \frac{1}{|A_s^i|}\exp(-2D/\epsilon)$. Correspondingly, the sum in \eqref{eq:limsum} is bounded from above by
\begin{equation}
\sum_{k\geq 0:s=s_k} \left(\frac{\alpha_{\#s}}{\smooth_{k}^i[a^i]}\right)^2 \leq |A_s^i|^2\exp\left(\frac{4D}{\epsilon}\right)\sum_{k\geq 0:s=s_k} \alpha_{\#s}^2.
\end{equation}
The right-hand side is a convergent sum by Assumption \ref{assume:first2}.

On the other hand, if Assumption \ref{assume:second} holds, then we no longer have a fixed lower bound on $\smooth_k^i[a^i]$. Instead, we have $\smooth_k^i[a^i]\geq \frac{1}{|A_s^i|}\exp\{-2D/\tau_{\#s}\}$. Correspondingly, the sum in \eqref{eq:limsum} is now bounded from above by
\begin{align}
\sum_{k\geq 0:s=s_k} \left(\frac{\alpha_{\#s}}{\smooth_{k}^i[a^i]}\right)^2 \leq |A_s^i|^2\sum_{k\geq 0:s=s_k} \alpha_{\#s}^2\exp\left(\frac{4D}{\tau_{\#s}}\right).
\end{align}
By Assumption \ref{assume:second2}, we have $\exp(4D/\tau_c)<C'\alpha_c^{-\rho}$ for all $c\geq C$. Therefore, we obtain
\begin{align}
\sum_{k\geq 0:s=s_k} \left(\frac{\alpha_{\#s}}{\smooth_{k}^i[a^i]}\right)^2 \leq |A_s^i|^2\sum_{c=0}^{C-1} \alpha_{c}^2\exp\left(\frac{4D}{\tau_{c}}\right) +  |A_s^i|^2C'\sum_{c\geq C} \alpha_{c}^{2-\rho}.
\end{align}
The right-hand side is a convergent sum by Assumption \ref{assume:second2}. This completes the proof.
\end{myproof}

\begin{lemma*}{Lemma \ref{lem:err}}
Suppose Assumption \ref{assume:common} and either Assumption \ref{assume:first} or \ref{assume:second} hold. Then, the error terms in \eqref{eq:discrete_main}, $\{\varepsilon^i_{s,k^{\dagger}}\}_{(i,s)\in\{1,2\}\times S}$ and $(\tau_{\#s+1}-\tau_{\#s})/\alpha_{\#s}$, are asymptotically negligible with probability $1$.
\end{lemma*}

\begin{myproof}
The error term $(\tau_{\#s+1}-\tau_{\#s})/\alpha_{\#s}$ is asymptotically negligible with probability $1$ either by Assumption \ref{assume:first2} or Assumption \ref{assume:second2}. On the other hand, the definition of $\varepsilon_{s',k^{\dagger}}^i$, as described in \eqref{eq:varepsilon}, and the evolution of $\hat{v}_{s',k}^i$ yield that
\begin{align}
\varepsilon_{s',k^{\dagger}}^i &= \frac{\sum_{l=k}^{k^{\dagger}-1}\mathbf{1}_{\{s_l=s'\}}\beta_{\#_ls'}(\inner{\hat{q}_{s',l}^i}{\smooth_{l}^i} - \hat{v}_{s',l}^i)}{\alpha_{\#_ks}},\quad \forall s'\in S.
\end{align}
Since $\{\beta_c\}_{c>0}$ is non-increasing by Assumption \ref{assume:common1} and the iterates are bounded by $D$ by Proposition \ref{prop:bound}, the error term $\varepsilon_{s',k^{\dagger}}^i$ is bounded from above by
\begin{equation}
|\varepsilon_{s',k^{\dagger}}^i| \leq \frac{\beta_{\#_ks'}}{\alpha_{\#_ks}}(k^{\dagger}-k)2D,\quad \forall s'\in S.
\end{equation}

By Borel-Cantelli Lemma, if we have
\begin{equation}\label{eq:Borel}
\sum_{k=0}^{\infty} \Prob{\frac{\beta_{\#_ks'}}{\alpha_{\#_ks}}(k^{\dagger}-k)>\lambda} < \infty
\end{equation}
for any $\lambda>0$, then $|\varepsilon_{s',k^{\dagger}}^i|\rightarrow 0$ as $k\rightarrow \infty$ with probability $1$. To this end, we will focus on the argument of the summation \eqref{eq:Borel}. Since $\#_ks\leq k$ and $\{\alpha_c\}_{c>0}$ is a non-increasing sequence by Assumption \ref{assume:common1}, we have
\begin{align}
&\Prob{\frac{\beta_{\#_ks'}}{\alpha_{\#_ks}}(k^{\dagger}-k)>\lambda} \leq\Prob{\frac{\beta_{\#_ks'}}{\alpha_{k}}(k^{\dagger}-k)>\lambda},\\
&\hspace{1in}= \sum_{\kappa=1}^{\infty}\Prob{k^{\dagger}-k=\kappa}\sum_{l=0:\frac{\beta_l}{\alpha_k}>\frac{\lambda}{\kappa}}^{k}\Prob{l=\#_ks'\,|\, k^{\dagger}-k=\kappa},\\
&\hspace{1in}= \sum_{\kappa=1}^{\infty}\Prob{k^{\dagger}-k=\kappa}\Prob{\#_ks'\leq \ell_k\left(\frac{\lambda}{\kappa}\right)\,|\, k^{\dagger}-k=\kappa},
\end{align}
where $\ell_k\left(\frac{\lambda}{\kappa}\right):=\max\left\{l\in\Xi_k\left(\frac{\lambda}{\kappa}\right)\right\}$ with $\Xi_k(\lambda):= \max\left\{l\in\mathbb{Z}\,:\, l\leq k \mbox{ and } \frac{\beta_l}{\alpha_k}> \lambda\right\}\cup\{-1\}$. Therefore, we obtain
\begin{align}\label{eq:mma}
\sum_{k=0}^{\infty} \Prob{\frac{\beta_{\#_ks'}}{\alpha_{\#_ks}}(k^{\dagger}-k)>\lambda} \leq \sum_{k=0}^{\infty} \sum_{\kappa=1}^{\infty}\Prob{k^{\dagger}-k=\kappa}\Prob{\#_ks'\leq \ell_k\left(\frac{\lambda}{\kappa}\right)\,|\, k^{\dagger}-k=\kappa}.
\end{align}
Since the argument of the summation on the right-hand side is non-negative, convergence of the following: 
\begin{align}
\sum_{\kappa=1}^{\infty}\sum_{k=0}^{\infty} \Prob{\kappa = k^{\dagger}-k}&\Prob{\#_ks'\leq \ell_k\left(\frac{\lambda}{\kappa}\right)\,|\, k^{\dagger}-k=\kappa}\nonumber\\
&=\sum_{\kappa=1}^{\infty}\Prob{\kappa = k^{\dagger}-k}\sum_{k=0}^{\infty} \Prob{\#_ks'\leq \ell_k\left(\frac{\lambda}{\kappa}\right)\,|\, k^{\dagger}-k=\kappa},\label{eq:mmma}
\end{align}
where the order of summations is interchanged would imply the convergence of \eqref{eq:mma}. Correspondingly, we will show that \eqref{eq:mmma} is convergent instead.

Partition the time axis into $n$-stage intervals and define $\bar{\#}_k^ns'$ as a counting process that increases by $1$ at the end of an $n$-stage interval if $s'$ is visited at least once within the last $n$-stage. More precisely, $\bar{\#}_k^ns'$ is, recursively, given by
\begin{align}
\bar{\#}_{k+1}^ns' = \left\{\begin{array}{ll} 
\bar{\#}_{k}^ns'+1 &\mbox{if } 0\equiv k\mod n \mbox{ and } \exists\, l \in (k-n,k]: s_l=s'\\
\bar{\#}_{k}^ns' & \mbox{o.w.}
\end{array}\right.
\end{align}
By its definition, we have $\bar{\#}_k^ns' \leq \#_ks'$, and therefore, we obtain
\begin{align}\label{eq:uppprob}
\Prob{\#_ks'\leq \ell_k\left(\frac{\lambda}{\kappa}\right)\,|\, k^{\dagger}-k=\kappa} \leq \Prob{\bar{\#}_k^ns'\leq \ell_k\left(\frac{\lambda}{\kappa}\right)\,|\, k^{\dagger}-k=\kappa}.
\end{align}
Either Assumption \ref{assume:first} or \ref{assume:second} yield that the probability that state $s'$ is visited within an $n$-stage interval is bounded from below, e.g., by some $\underline{p}>0$, for every sequence of actions. Correspondingly, 
Correspondingly, the probability that $s'$ is not visited within the $n$-stage interval is bounded from above by $1-\underline{p}$.
Therefore, we can bound the right-hand side of \eqref{eq:uppprob} from above by
\begin{align}\label{eq:uppbinom}
\Prob{\bar{\#}_k^ns'\leq \ell_k\left(\frac{\lambda}{\kappa}\right)\,|\, k^{\dagger}-k=\kappa} \leq \sum_{l=0: l\leq \ell_k\left(\frac{\lambda}{\kappa}\right)}^{\left\lfloor \frac{k}{n}\right\rfloor} \binom{\left\lfloor \frac{k}{n}\right\rfloor}{l} 1^l(1-\underline{p})^{\left\lfloor \frac{k}{n}\right\rfloor-l}.
\end{align}

Assumption \ref{assume:common2} yields that for any $M\in(0,1)$, there exists a non-decreasing polynomial function $C_M(\cdot)$ such that 
\begin{equation}\label{eq:ell}
\ell_k \left(\frac{\lambda}{\kappa}\right) \leq M k < nM\left(\left\lfloor \frac{k}{n}\right\rfloor + 1\right), \quad \forall k \geq C_M\left(\frac{\kappa}{\lambda}\right).
\end{equation}
For $k\geq n$, \eqref{eq:ell} can also be written as
\begin{equation}
\frac{\ell_k\left(\lambda/\kappa\right)}{\left\lfloor k/n\right\rfloor}\leq nM\left(1+\frac{1}{\lfloor k/n \rfloor}\right)\leq 2nM,\quad \forall l\geq \max\{C_M(\kappa/\lambda),n\}.
\end{equation}
Since $M\in(0,1)$ is arbitrary, there exists $M$ such that $2nM<1$, i.e, $M\in(0,1/2n)$. Therefore, we have
\begin{equation}
\sum_{l=0: l\leq \ell_k\left(\frac{\lambda}{\kappa}\right)}^{\left\lfloor \frac{k}{n}\right\rfloor} \binom{\left\lfloor \frac{k}{n}\right\rfloor}{l} 1^l(1-\underline{p})^{\left\lfloor \frac{k}{n}\right\rfloor-l} 
\leq (1-\underline{p})^{\left\lfloor \frac{k}{n}\right\rfloor - \ell_k\left(\frac{\lambda}{\kappa}\right)}\sum_{l=0}^{\ell_k\left(\frac{\lambda}{\kappa}\right)} \binom{\left\lfloor \frac{k}{n} \right\rfloor}{l},\label{eq:nm}
\end{equation}
for all $k\geq \max\{C_M(\kappa/\lambda),n\}$, since $(1-\underline{p})\in[0,1)$. Furthermore, we can set $M\in(0,1/2n)$ such that $2nM<1/2$. Then, we can resort to the following inequality \cite[Lemma 16.19]{ref:Flum06book}
\begin{equation}
\sum_{l=0}^{L} \binom{k}{l}\leq 2^{H_2\left(L/k\right)k},\quad\mbox{if } L/k\leq 1/2,
\end{equation} 
where $H_2(p):=-p\log(p)-(1-p)\log(1-p)$. Therefore, for all $k\geq \max\{C_M(\kappa/\lambda),n\}$, we obtain
\begin{align}
(1-\underline{p})^{\left\lfloor \frac{k}{n}\right\rfloor - \ell_k\left(\frac{\lambda}{\kappa}\right)}\sum_{l=0}^{\ell_k\left(\frac{\lambda}{\kappa}\right)} \binom{\left\lfloor \frac{k}{n}\right\rfloor}{l} 
&\leq (1-\underline{p})^{\left\lfloor \frac{k}{n}\right\rfloor - \ell_k\left(\frac{\lambda}{\kappa}\right)}2^{H_2\left( \frac{\left\lfloor\ell_k\left(\lambda/\kappa\right)\right\rfloor}{\left\lfloor k/n\right\rfloor}\right)\left\lfloor k/n\right\rfloor}\\
&\leq (1-\underline{p})^{\left\lfloor \frac{k}{n}\right\rfloor - 2nM\left\lfloor \frac{k}{n}\right\rfloor}2^{H_2(2nM)\left\lfloor \frac{k}{n}\right\rfloor}\\
&=\left[(1-\underline{p})^{1 - 2nM}2^{H_2(2nM)}\right]^{\left\lfloor \frac{k}{n}\right\rfloor},\label{eq:eta}
\end{align}
since $(1-\underline{p})\in[0,1)$ and $H_2(p)$ is an increasing function for $p\in(0,1/2)$.

We define $\eta_M:=(1-\underline{p})^{1 - 2nM}2^{H_2(2nM)}$. Note that for $M=0$, we have $\eta_0 = (1-\underline{p})2^{H_2(0)}=(1-\underline{p})\in[0,1)$. By the continuity of $\eta_M$ in $M$, there exists $M\in(0,1/4n)$ such that $\eta_M\in(0,1)$. By \eqref{eq:uppprob}, \eqref{eq:uppbinom}, \eqref{eq:nm}, and \eqref{eq:eta}, we obtain
\begin{align}
\Prob{\#_ks'\leq \ell_k\left(\frac{\lambda}{\kappa}\right)\,|\, k^{\dagger}-k=\kappa} \leq \eta_M^{\left\lfloor\frac{k}{n}\right\rfloor},\quad \forall k\geq \max\left\{C_M\left(\frac{\kappa}{\lambda}\right),n\right\}.
\end{align}
Its sum over $k\geq 0$ (corresponding to the inner sum in \eqref{eq:mmma}) is bounded from above by
\begin{align}
\sum_{k=0}^{\infty}\Prob{\#_ks'\leq \ell_k\left(\frac{\lambda}{\kappa}\right)\,|\, k^{\dagger}-k=\kappa}
&\stackrel{(a)}{\leq}  C_M\left(\frac{\kappa}{\lambda}\right) + n + \sum_{k=0}^{\infty}\eta_M^{\left\lfloor \frac{k}{n}\right\rfloor}\\
&\stackrel{(b)}{\leq}  C_M\left(\frac{\kappa}{\lambda}\right) + n + \sum_{k=0}^{\infty}\eta_M^{\frac{k}{n}-1}\\
&\stackrel{(c)}{=}  C_M\left(\frac{\kappa}{\lambda}\right) + n + \frac{1}{\eta_M(1-\eta_M^{1/n})},\label{eq:innerbound}
\end{align}
where $(a)$ follows since any probability is bounded from above by one and $\max\{C_M(\kappa/\lambda),n\}\leq C_M(\kappa/\lambda)+n$; $(b)$ follows since $\lfloor k/n\rfloor > k/n-1$ and $\eta_M\in(0,1)$; and $(c)$ follows since $\eta_M^{1/n}\in(0,1)$ when $n>0$.

Based on \eqref{eq:innerbound}, the sum \eqref{eq:mmma} is bounded from above by
\begin{align}
&\sum_{\kappa=1}^{\infty} \Prob{k^{\dagger}-k=\kappa}\sum_{k=0}^{\infty}\Prob{\#_ks'\leq \ell_k\left(\frac{\lambda}{\kappa}\right)\,|\, k^{\dagger}-k=\kappa}\nonumber \\
&\hspace{.5in}\leq \sum_{\kappa=1}^{\infty} \Prob{k^{\dagger}-k=\kappa}\left(C_M\left(\frac{\kappa}{\lambda}\right) + n + \frac{1}{\eta_M(1-\eta_M^{1/n})}\right)\\
&\hspace{.5in}\leq \sum_{t=1}^{\infty} \Prob{(t-1)n<k^{\dagger}-k\leq tn}\left(C_M\left(\frac{tn}{\lambda}\right) + n + \frac{1}{\eta_M(1-\eta_M^{1/n})}\right),\label{eq:fin}
\end{align}
where the last inequality follows since $C_M(\cdot)$ is non-decreasing. Either Assumption \ref{assume:first} or \ref{assume:second} yield that
\begin{equation}\label{eq:righthandside}
 \Prob{(t-1)n<k^{\dagger}-k\leq tn} \leq (1-\underline{p})^{t-1}.
\end{equation}
Therefore, we can bound \eqref{eq:fin} from above by
\begin{align}
\sum_{t=1}^{\infty}(1-\underline{p})^{t-1}\left(C_M\left(\frac{tn}{\lambda}\right) + n + \frac{1}{\eta_M(1-\eta_M^{1/n})}\right).
\end{align}
Since $C_M(\cdot)$ is a polynomial function by Assumption \ref{assume:common2}, the ratio test, i.e., 
\begin{equation}
\lim_{t\rightarrow\infty}\frac{(1-\underline{p})^{t}\left(C_M\left(\frac{(t+1)n}{\lambda}\right) + n + \frac{1}{\eta_M(1-\eta_M^{1/n})}\right)}{(1-\underline{p})^{t-1}\left(C_M\left(\frac{tn}{\lambda}\right) + n + \frac{1}{\eta_M(1-\eta_M^{1/n})}\right)}=1-\underline{p}<1,
\end{equation}
yields that \eqref{eq:righthandside} is convergent for any $\lambda >0$. Therefore, we obtain \eqref{eq:Borel}, which completes the proof. 
\end{myproof}

\begin{lemma*}{Lemma \ref{lem:Lyapunov}}
Consider any trajecttory of \eqref{eq:cont_q_higher} and let $x(t):=(q^1(t),q^2(t),\pi^1(t),\pi^2(t))$. Then the candidate function $V(\cdot)$, as described in \eqref{eq:Lyapunov}, satisfies 
\begin{itemize}
\item $V(x(t')) < V(x(t))$ for all $t'>t$ if $V(x(t))>0$,
\item $V(x(t')) = 0$ for all $t'>t$ if $V(x(t))=0$.
\end{itemize} 
\end{lemma*}

\begin{myproof}
Fix an arbitrary solution $(q^1(t),q^2(t),\pi^1(t),\pi^2(t))$ to the o.d.e. \eqref{eq:cont_q_higher} for some arbitrary initial point, and define the functions $L:[0,\infty)\rightarrow \mathbb{R}$ and $H:[0,\infty)\rightarrow\mathbb{R}$ by
\begin{align}
&L(t):= \sum_{i=1,2}\max_{\mu\in\Delta(A^i)}\{\inner{\mu}{q^i(t)}+\tau\nu^i(\mu)\} - \lambda \zeta\label{eq:L}\\
&H(t):= \|q^1(t)-Q^1\pi^2(t)\|^2 + \|q^2(t)-Q^2\pi^1(t)\|^2.
\end{align} 
Then, we obtain
\begin{equation}\label{eq:def}
V(q^1(t),q^2(t),\pi^1(t),\pi^2(t)) = \left[L(t)\right]_+ + H(t).
\end{equation}

Note that $H(t)$ is a continuous, differentiable and non-negative function. Its time derivative is given by
\begin{align}\label{eq:dHdt}
\frac{dH(t)}{dt} = - 2 H(t).
\end{align}
If $\zeta=0$, then $L(\cdot)$ reduces to the Lyapunov function introduced by \cite{ref:Harris98} for continous-time best response dynamics in zero-sum strategic-form games, and therefore, $V(\cdot)$ is a Lyapunov function function for \eqref{eq:cont_q_higher}. 

Suppose that $\zeta>0$. Note that $L(\cdot)$ is also a continuous and differentiable function, but it can also be negative (when $\zeta>0$) and the last term of \eqref{eq:L}, i.e., $-\lambda\zeta$, is time-invariant. For notational simplicity, let $\smooth^i(t):=\smoothFunc^i(q^i(t),\tau)$. Based on the envelope theorem (which can be invoked due to the smoothness of the optimization argument induced by $\nu^i(\cdot)$), the time derivative of $L(t)$ is given by
\begin{align}
\frac{dL(t)}{dt} =& \sum_{i=1,2}\inner{\smooth^i(t)}{\frac{dq^i(t)}{dt}}\\
=& \sum_{i=1,2}\inner{\smooth^i(t)}{\left(Q^i\smooth^{-i}(t)-q^i(t)\right)}\\
=& - \sum_{i=1,2} \left(\inner{\smooth^i(t)}{q^i(t)} + \tau\nu^i(\smooth^i(t))\right) \nonumber\\
&+ \smooth^1(t)^T\left(Q^1+(Q^2)^T\right)\smooth^2(t) + \tau\sum_{i=1,2}\nu^i(\smooth^i(t)).
\end{align}
By definition, $\smooth^i(t)\in\Delta(A^i)$ and $\nu^i(\smooth^i(t))\leq \log(|A^i|)$. Therefore, we have
\begin{equation}
\smooth^1(t)^T\left(Q^1+(Q^2)^T\right)\smooth^2(t) + \tau\sum_{i=1,2}\nu^i(\smooth^i(t)) \leq \zeta,
\end{equation}
where $\zeta$ is as described in \eqref{eq:zeta}. Correspondingly, the time derivative of $L(t)$ is bounded from above by
\begin{align}
\frac{dL(t)}{dt} &< - \sum_{i=1,2} \left(\inner{\smooth^i(t)}{q^i(t)} + \tau\nu^i(\smooth^i(t))\right) + \lambda \zeta\\
&= -L(t),\label{eq:dLdt}
\end{align}
where the strict inequality follows since $\lambda > 1$ and $\zeta>0$. This yields that $L(t)$ is {\em strictly} decreasing whenever $L(t)\geq 0$. Therefore, $\{q^1(t),q^2(t),\pi^1(t),\pi^2(t):L(t) \leq 0\}$ is a positively invariant set for any trajectory. In other words, if $L(t') \leq 0$ for some $t'$, then $L(t'')\leq0$ for all $t''>t'$. Therefore, by \eqref{eq:def}, the time-derivatives \eqref{eq:dHdt} and \eqref{eq:dLdt} yield that $V(\cdot)$ is a Lyapunov function, which completes the proof.
\end{myproof}

\section{Proof of Corollary \ref{cor:rational} to Theorem \ref{thm:main}}\label{sec:cors}

\begin{lemma*}{Corollary \ref{cor:rational}}
Suppose that player $-i$ follows an (asymptotically) stationary strategy 
$\{\tilde{\pi}_s^{-i}\in \mathrm{int}\Delta(A_s^i)\}_{s\in S}$ while player $i$ adopts the learning dynamics described in Table \ref{algo}, and Assumption \ref{assume:common} holds. 
Then, the asymptotic behavior of the value function estimate $\{\hat{v}_{s,k}^i\}_{k\geq 0}$ is given by 
\begin{subequations}
\begin{align}
&\limsup_{k\rightarrow\infty}\left|\hat{v}_{s,k}^i - \max_{\pi^i}~v_{\pi^i,\tilde{\pi}^{-i}}^i(s)
\right| \leq \epsilon \xi^i g(\gamma),\quad \mbox{under Assumption \ref{assume:first}},\\
&\lim_{k\rightarrow\infty} \left|\hat{v}_{s,k}^i - \max_{\pi^i}~v_{\pi^i,\tilde{\pi}^{-i}}^i(s)
\right| = 0,\hspace{.5in}\quad \mbox{under Assumption \ref{assume:second}}
\end{align}
\end{subequations}
for all $s\in S$, w.p. $1$, where $\xi^i:= \max_{s'\in S}\left\{\log(|A_{s'}^i|)\right\}$ and $g(\cdot)$ is as described in Theorem \ref{thm:main}. 

Furthermore, the asymptotic behavior of the weighted averages $\{\hat{\pi}_{k}^i\}_{k\geq 0}$, described in Theorem \ref{thm:main}, is given by
\begin{subequations}\label{eq:result2}
\begin{align}
&\limsup_{k\rightarrow\infty}\Big(\max_{\pi^i}~v_{\pi^i,\tilde{\pi}^{-i}}^i(s) - v_{\hat{\pi}_{k}^{i},\tilde{\pi}^{-i}}^i(s)
\Big)
\leq \epsilon \xi^i h(\gamma),\quad\hspace{-.05in} \mbox{under Assumption \ref{assume:first}},\\
&\lim_{k\rightarrow\infty}\Big(\max_{\pi^i}~v_{\pi^i,\tilde{\pi}^{-i}}^i(s) - v_{\hat{\pi}_{k}^{i},\tilde{\pi}^{-i}}^i(s)\Big)
= 0,\hspace{.58in}\mbox{under Assumption \ref{assume:second}},\end{align}
\end{subequations}
for all $s\in S$, w.p. $1$, where $h(\gamma)$ is as described in Theorem \ref{thm:main}, i.e., these weighted-average strategies  converge to near or exact best-response strategy, depending on whether Assumption \ref{assume:first} or \ref{assume:second} hold.
\end{lemma*}

\begin{myproof}
The proof follows from the observation that Theorem \ref{thm:main} can be generalized to the scenarios where nature draws $r_s^i(a^1,a^2)\in [-D,D]$ and $p(s'|s,a^1,a^2)$ depending on a random event in a rather straightforward way since it only introduces a stochastic approximation error that is a square integrable Martingale difference sequence. For example, player $i$ receives $r_s^i(\omega_k,a^1,a^2)$ with random event $\omega_k\in \Omega$ and $r_s^i(\omega_k,a^1,a^2)\in [-D,D]$ for all $\omega\in \Omega$ and state transitions are governed by the kernel $p(s'|s,\omega_k,a^1,a^2)$ while $\omega_k\rightarrow\omega_o$ as $k\rightarrow \infty$ with probability $1$. 

Since player $-i$ follows an asymptotically stationary strategy $\{\tilde{\pi}^{-i}_s\}_{s\in S}$ almost surely, we can view player $-i$ as nature and its action at stage $k$, $a^{-i}_k$, as the random event $\omega_k$. Then, it reduces into a single-player game. We can still invoke Theorem \ref{thm:main} if we introduce an auxiliary player $i'$ that has a single action at every state without loss of generality. This completes the proof.  
\end{myproof}

\section{Additional Simulation Setup}\label{sec:add_sim}

We consider a larger-scale case with $|S|=20$ states and $|A^i_s|=10$ actions per state. The discount factor  $\gamma=0.5$. The reward functions are chosen randomly in a way that $r^1_s(a^1,a^2)\propto \bar{r}_{s,a^1,a^2}$ for $s\in S$, where $\bar{r}_{s,a^1,a^2}$ is uniformly drawn from $[-1,1]$. Then, $r^1_s(a^1,a^2)$ is normalized by $\max_{s,a^1,a^2}\{r^1_s(a^1,a^2)\}/2$ so that $|r_s^i(a^1,a^2)|\leq R=2$ for all $(i,s,a^1,a^2)$. For the state transition dynamics $p$, we construct two cases, \textbf{Case 3} and \textbf{Case 4} by randomly generating transition probabilities, in a way that they  satisfy Assumptions  \ref{assume:firstIO} and \ref{assume:secondIO}, respectively. For \textbf{Case 3} and \textbf{Case 4}, we  choose  the temperature parameter as $\tau_c=\max\{\epsilon,\tau_c'\}$ and as $\tau_c'$ in \eqref{equ:tau_to_zero}, respectively, with $\epsilon=2\times 10^{-2}$, $\bar{\tau}=0.1$. For both cases, we choose the stepsizes $\alpha_c=1/c^{0.9}$ and $\beta_c=1/c$ with $\rho_\alpha=0.9$, $\rho_\beta=1$, and $\rho=0.85$ for the $\tau_c'$ in \eqref{equ:tau_to_zero}.    The simulation results are illustrated in Figure \ref{fig:simulation_results_2}. Note that as the number of states is large, the plot becomes very dense and cluttered if both players' curves are plotted, together with the standard-deviation bar-area, as in Figure \ref{fig:simulation_results_2}. We thus only plot an example trial, with only Player $2$'s curves, and the summation of the value function estimates. The convergence of Player $1$'s value function estimates can be deduced accordingly. It is seen from Figure \ref{fig:simulation_results_2} that our theory can be corroborated by simulations even for this larger-scale case.

\end{document}